\documentclass[twoside]{cernrep} 
\usepackage{texnames}
\usepackage{amsthm}
\usepackage{amsmath}
\usepackage[T1]{fontenc}
\usepackage[bookmarks, colorlinks=true, linktoc=page, pdftex, linkcolor=blue, citecolor=blue, urlcolor=blue]{hyperref}

\usepackage{lipsum}

\newcommand\blfootnote[1]{%
  \begingroup
  \renewcommand\thefootnote{}\footnote{#1}%
  \addtocounter{footnote}{-1}%
  \endgroup
}

\begin{document}
 \setcounter{page}{149} 
\title{Practical statistics for particle physics}

\author {R.~J.~Barlow}

\institute{The University of Huddersfield, Huddersfield, United Kingdom}

\begin{abstract}
This is the write-up of a set of 
lectures given at the Asia Europe Pacific School of High Energy Physics in Quy Nhon, Vietnam in September 2018,
to an audience of PhD students in all branches of particle physics. They cover the different meanings of `probability', particularly  frequentist and Bayesian, the binomial, the Poisson and the Gaussian distributions, hypothesis testing, estimation, errors (including asymmetric and systematic errors) and goodness of fit.
Several  different methods used in setting upper limits are explained, followed by a 
discussion on why 5 sigma are conventionally required for a `discovery'.
 \end{abstract}


\maketitle 

 \def \etal {{\it et al.}}
 \def \half {{1 \over 2}}

\def \ifb {${\rm fb}^{-1}$}


\section{Introduction}
\blfootnote{\textcopyright~CERN, 2020, \href{https://creativecommons.org/licenses/by/4.0/}{CC-BY-4.0 licence},
\href{http://doi.org/10.23730/CYRSP-2020-005.149}{doi:10.23730/CYRSP-2020-005.149}, ISSN 0531-4283.}  
To interpret the  results of your particle physics experiment and see what it implies for the relevant theoretical model and parameters, you need to use statistical techniques. These are a part of your experimental toolkit, and to extract the maximum information from your data you need to use the correct and most powerful statistical tools. 

Particle physics (like, probably, any field of science) has is own special set of statistical processes and language. 
Our use is in some ways more complicated (we often fit multi-parameter functions, not just straight lines) and in some
ways more simple (we do not have to worry about ethics, or law suits). 
So the generic textbooks and courses you will meet on `Statistics' are not really appropriate. That's why HEP schools like this one include lectures on statistics as well as the fundamental real physics, like field theory and physics beyond the Standard Model (BSM). 
 
 There are several textbooks~\cite{Barlow, Cowan, Narsky, Behnke, Lyons, Zech} available which are designed for an audience of particle physicists.
 You will find these helpful---more helpful than general statistical  textbooks.  You should find one 
 whose language suits you and keep a copy on your bookshelf---preferably purchased---but at least  on long term library loan. You will also find useful conference proceedings~\cite{Phystat1, Phystat2, Phystat3}, journal papers (particularly in Nuclear Instruments and Methods) and web material: often your own experiment will have a set of pages devoted to the topic. 
 
\section{Probability}

We begin by looking at the concept of probability. Although this is familiar (we use it all the time, both 
inside and outside the laboratory), its use is not as obvious as you would think. 

 \subsection {What is probability?}

 A typical exam for Statistics101 (or equivalent) might well contain the question:
 
 \
 
 \hrule
 
 \
 
 {\rm \qquad Q1 \qquad Explain what is meant by the {\it probability $P_A$} of an event {\it A} \hfil
 \ \vbox {\hsize  6 cm  }\qquad \qquad \hfill [1] 
 }
 
 \
 
 \hrule
 
 \
 
 The `1' in square brackets signifies that the answer carries one mark. That's an indication that just a sentence or two are required, not a long essay.
 
 Asking a group of physicists this question produces answers falling into four different categories

 \begin{enumerate}
 \item $P_A$ is  number obeying certain mathematical rules,

 \item $P_A$ is a property of $A$ that determines how often $A$ happens,

 \item For $N$ trials in which $A$ occurs $N_A$ times, $P_A$ is the limit of $N_A / N$ for large $N$, and

 \item $P_A$ is my belief that $A$ will happen, measurable by seeing what odds I will accept in a bet.
 \end{enumerate}
 
Although all these are generally present, number 3 is the most common, perhaps because it is 
often explicitly taught as the definition. All are, in some way, correct!
We consider each in turn. 

\subsection {Mathematical probability}

The Kolmogorov axioms are: 
 For  all $A \subset S$

\begin{equation}
\begin{split}
P_A\geq 0 \\
P_S=1\\
P_{A \cup B} =P_A+P_B \  {\rm if}\  A \cap B =\phi \ {\rm and}\  A,B \subset S
\end{split}
\quad.
\end{equation}

From these simple axioms a complete and complicated structure of theorems can be erected.
This is what pure mathematicians do.  For example,
the 2nd and 3rd axiom show that the probability of not-$A$ $P_{\overline A}$, is $1-P_A$, and 
then the 1st axiom shows that 
$P_A \leq 1$: probabilities must be less than 1.  

 But
the axioms and the ensuing theorems says nothing about what $P_A$ actually means.  
Kolmogorov had frequentist probability in mind, but these axioms apply to any definition: he explicitly avoids
tying $P_A$ down in this way. So although this apparatus enables us to compute numbers, it does not tell us
what we can use them for.

\subsection{ Real probability}

Also known as Classical probability, this was developed during the 18th--19th centuries 
by Pascal, Laplace and others to serve the gambling industry.

If there are several possible outcomes and there is a symmetry between them so they are all, in a sense, identical,
then their individual probabilities must be equal. For example, there are 
two sides to a coin, so if you toss it there must be a  probability $\frac{1}{2}$ for each face to land uppermost.
Likewise there are 52 cards in a pack,  so the probability of a particular card being chosen is $1 \over 52$.  
 In the same way there are 6 sides to a dice, and 33 slots in a roulette wheel.

 This enables you to 
answer questions like `What is the probability of rolling more than 10 with 2 dice?'.  There are 3 such combinations (5-6, 6-5 and 6-6) out of the $6 \times 6 = 36$ total possibilities, so the probability is $1 \over 12$. Compound instances of $A$ are broken down into smaller instances to which the symmetry argument can be applied. This is satisfactory and clearly applicable---you know that if someone offers you a 10 to 1 bet on this dice throw, you should refuse; in the long run
knowledge of the correct probabilities will pay off.

 The problem arises that this approach
 cannot be applied to continuous variables.  This is brought out in Bertan's paradoxes, one of which runs:
 
 {\it In a circle of radius $R$ an equilateral triangle is drawn.
 A chord is drawn at random.  What is the probability that the length of the chord is greater than the side
 of the triangle?}
 
 \begin{figure}[h]
 \centerline{
 \includegraphics[width=8 cm ]{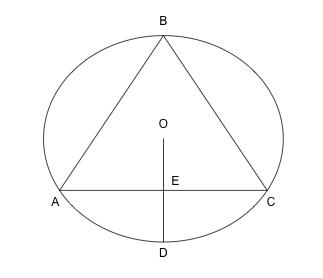}
 }
 \caption{\label{fig:bertan} Bertan's paradox}
 \end{figure}
 
 Considering Fig.~\ref{fig:bertan} one can give three answers:
 
 \begin{enumerate}
 \item If the chord, without loss of generality, starts at A, then it will be longer than the side if the end point is anywhere between B and C. So the answer is obviously $1 \over 3$.
 \item  If the centre of the chord, without loss of generality, is chosen at random along the line OD, then it will be longer than the side of the triangle if it is in OE rather than ED. E is the midpoint of OD so the answer is obviously $1 \over 2$.
 \item  If the centre of the chord, without loss of generality, is chosen at random within the circle, then it will be longer than the side of the triangle if it lies within the circle of radius $R \over 2$. So the answer is obviously $1 \over 4$.
 \end{enumerate}
  
  So we have three obvious but contradictory answers. The whole question is built on a false premise: drawing a chord `at random' is, unlike tossing a coin or throwing a dice, not defined. 
  Another way of seeing this is that a distribution which is uniform in one variable, say $\theta$, is not uniform in any non-trivial transformation of that variable, say 
 $\cos \theta$ or $\tan \theta$.  Classical probability has therefore to be discarded.

\subsection{Frequentist probability}

Because of such difficulties,  Real Probability was replaced by Frequentist Probability in the early 20th century.
This is the usual definition taught in schools and undergraduate classes. A very readable account is given by von Mises~\cite{vonMises}:

$$P_A = \lim_{N \to \infty} {N_A \over N}\quad.$$

$N$ is the total number of events in the ensemble (or collective). It can be visualised as a Venn diagram, as in Fig.~\ref{fig:venn}.

 \begin{figure}[h]
 \centerline{
 \includegraphics[width=8 cm ]{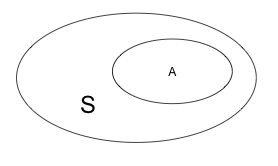}
 }
 \caption{\label{fig:venn} Frequentist probability}
 \end{figure}

The probability of a coin landing heads up is $1 \over 2$ because if you toss a coin $1000$ times, one side will come down $\sim 500$ times. That is an empirical definition (Frequentist probability has roots in the Vienna school and logical positivism). 
Similarly, 
the lifetime of a muon is $2.2 \mu$s because if you take 1000 muons and wait $2.2 \mu$s,  then $\sim 368 $ (that's a fraction $e^{-1}$)  will remain. 

With this definition
 $P_A$ is not just a property of $A$ but a joint property of $A$ and the ensemble. 
 The same coin will have a different probability for showing head depending on whether it is in a purse or in a numismatic collection. This leads to two distinctive properties (or, some would say, problems) for frequentist probability.

 Firstly, there may be 
 more than one ensemble.  To take an everyday example from von Mises,
German life insurance companies  pay out on 0.4\% of 40 year old male clients. Your friend  Hans is 40 today. What is the probability that he will survive to see his 41st birthday?
99.6\% is an answer (if he's insured).
But he is also a non-smoker and non-drinker---so perhaps the figure is higher (maybe 99.8\%)?
But if he  drives a Harley-Davidson it should be lower (maybe 99.0\%)?
All these numbers are acceptable. The individual Hans belongs to several different ensembles, and the probability will be different for each of them.

 To take an example from physics, suppose your experiment has a Particle Identification (PID) system using Cherenkov, time-of-flight and/or $dE\over dx$ measurements. You want to talk about  
the probability that a $K^+$ will be correctly  recognised by your PID.
You determine this by 
considering many  $K^+$ mesons  and counting the number accepted to get $P=N_{acc}/N_{tot}$.
But these will depend on the kaon sample you work with. It could be all kaons, or kaons above a certain energy threshold, or that actually enter the detector. The ensemble can be defined in various ways, each giving a valid but different 
value for the probability.

On the other hand, there may be no Ensemble.
To take an everyday example we might want to 
calculate
 the probability that it will rain tomorrow. This is impossible.   There is only one tomorrow.  It will either rain or not rain.  $P_\mathrm{rain}$ is either 0 or 1, and we won't know which until tomorrow gets here. Von Mises insists that 
 statements like `It will probably rain tomorrow' are loose and unscientific.

To take an example from physics, consider the
 probability  that there is a supersymmetric particle with mass below 2 TeV. Again, either there is or there isn't. 
  
  But, despite von Mises' objections, it does seem sensible, as the pressure falls and the gathering clouds turn grey,
  to say `It will probably rain'. So this is a drawback to the frequentist definition.  
  We will return to this and show how frequentists can talk meaningfully and quantitatively about
  unique events in the discussion of confidence intervals in Section~\ref{sec:confidence}.

\subsection{Bayes' theorem}

Before presenting Bayesian statistics we need to discuss 
Bayes' theorem, though we point out that 
Bayes' theorem applies (and is useful)  in any probability model: it goes right back to the Kolmogorov axioms. 

First we need to define the
conditional probability: $P(A|B)$: this is the probability for $A$, given that $B$ is true. For
example:  if a playing card is drawn at random from a pack of 52, then 
$P(\spadesuit A) = {1 \over 52}$,
but if you are told that the card is black, then 
$P(\spadesuit A|Black) = {1 \over 26}$ (and obviously $P(\spadesuit A|Red)=0$).

Bayes' theorem is just
\begin{equation}
P(A|B)= {P(B|A) \over P(B) } \times P(A) \quad.
\end{equation}

The proof is gratifyingly simple: the probability that $A$ and $B$ are both true can be written in two ways

\centerline{$P(A|B) \times  P(B) = P(A \& B) =  P(B|A) \times P(A)$\quad.}

Throw away middle term and divide by $P(B)$ to get the result.

As a first example, we go back to the ace of spades above. A card is drawn at random, and you are told that it is black. Bayes' theorem says
 
$P(\spadesuit A|Black) =  {P(Black | \spadesuit A) \over P(Black)} P(\spadesuit A) = {1  \over {1 \over 2}} \times {1 \over 52 }= {1 \over 26}$\quad; 

\noindent i.e. the original probability of drawing $\spadesuit A$,  $1 \over 52$, is multiplied by the probability that the ace of spades is black (just 1) and divided by the overall probability of drawing a black card ($1 \over 2$) to give the obvious result.
 
 For a less trivial example, suppose you have a momentum-selected  beam which
 is 90\% $\pi$ and 10\% $K$. This goes through a Cherenkov counter
 for which pions exceed the threshold velocity but kaons do not. 
 In principle pions will give a signal, but suppose there is a 5\% chance, due to inefficiencies,  that they will not. 
 Again in principle kaons always give no Cherenkov signal, but suppose 
 that probability is only   95\% due to background noise.  What is the probability that a particle 
 identified as a kaon, as it gave no signal, is truly one?
 
 Bayes' theorem runs
 
$P(K|no\ signal)= {P(no\ signal|K) \over P(no\ signal)} \times P(K)={0.95 \over 0.95 \times 0.1 + 0.05 \times 0.9} \times 0.1 = 0.68 \quad,$ 

\noindent showing that the probability is only $2 \over 3$. The positive  identification is not enough to overwhelm the 9:1 $\pi:K$ ratio. Incidentally this uses the (often handy) expression for the denominator: $P(B)=P(B|A) \times P(A) + P(B|\overline A)\times  \overline{P(A)}$.

 \subsection {Bayesian probability}
 
 The Bayesian definition of probability is that $P_A$ represents your belief in $A$.
1 represents certainty, 0 represents total disbelief.
Intermediate values can be calibrated by asking whether you would prefer to bet on $A$, or on a white ball being drawn from an urn containing a mix of white and black balls. 

This avoids the limitations of frequentist probability---coins, dice, kaons, rain tomorrow, existence of supersymmetry (SUSY) can all have probabilities assigned to them.

 The drawback is that your value for $P_A$ may be different from mine, or anyone else's. It is
 also called subjective probability.
 
 Bayesian probability makes great use of  Bayes' theorem, in the form

\begin{equation}P(Theory|Data)= {P(Data|Theory) \over P(Data) } \times P(Theory) \quad.
\end{equation}

$P(Theory)$ is called the {\em prior}: your initial belief in $Theory$. $P(Data|Theory)$  is the {\em Likelihood}:
the probability of getting $Data$ if $Theory$ is true. $P(Theory|Data)$ is the {\em Posterior}: your belief in $Theory$ 
in the light of a particular $Data$ being observed.

So this all works very sensibly.  If the data observed is predicted by the theory, your belief in that theory is boosted,
though this is moderated by the probabilty that the data could have arisen anyway. Conversely, if data is observed which 
is disfavoured by the theory, your belief in that theory is weakened.

The process can be chained. 
The posterior from a first experiment can be taken as the prior for a second experiment, and so on. 
When you write out the factors you find that the order doesn't matter. 
 
 \subsubsection{Prior distributions}
 
Often, though, the theory being considered is not totally defined: it may contain a parameter (or several parameters)
 such as a mass, coupling constant, or decay rate.  Generically we will call this $a$, with the proviso that it may
 be multidimensional.  

The prior is now not a single number $P(Theory)$ 
but a probability distribution $P_0(a)$. 
$\int_{a_1}^{a_2} P_0(a)\, da $ is your prior belief that $a$ lies between $a_1$ and $a_2$.
$\int_{-\infty}^{\infty} P_0(a)\, da$ is your original $P(Theory)$. This is generally taken as 1, which is valid provided the possibility that the theory that is false is matched by some value of $a$---for example if the coupling constant for a hypothetical particle is zero, that accommodates any belief that it might not exist. Bayes' theorem then runs:

\begin{equation} P_1(a;x) \propto L(a;x) P_0(a)
\label{eq:bayes}
\quad.
\end{equation}

If the  range of $a$ is infinite, $P_0(a)$ may be vanishingly small (this is called an `improper prior'). 
However this is not a problem.  Suppose, for example, that all we know about $a$ is that it is non-negative, and we
are genuinely equally open to its having any value. We write $P_0(a)$ as $C$, so $\int_{a_1}^{a_2} P_0(a)\, da =C(a_2-a_1)$.
This probability is vanishingly small: if you were offered the choice of a bet on $a$ lying within the range $[a_1,a_2]$
or of drawing a white ball from an urn containing 1 white ball and $N$ black balls, you would choose the latter, however large $N$ was.  However it is not zero: if the  urn contained $N$ black balls, but no white ball, your betting choice would change.  After a measurement you have
  $P_1(a;x)={L(a;x) \over \int L(a';x) C da'} C$, and the factors of $C$ can be cancelled (which, and this is the point, you could {\em not} do if $C$ were exactly zero) giving 
  $P_1(a;x)={L(a;x) \over \int L(a';x)  da'} $ or,
  $
  P_1(a;x) \propto L(a;x) $,
 and you can then just normalize $P_1(a)$ to 1.

 \begin{figure}
 [h]
 \centerline{
\includegraphics [width=4.8 cm,angle=270,trim={0 0 0 0},clip] 
{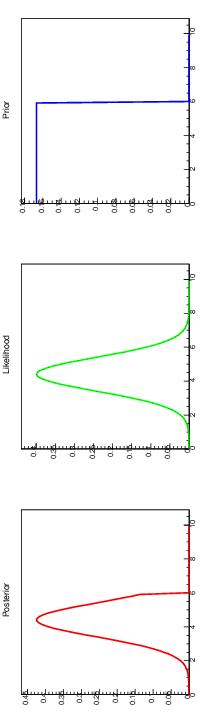}}
\caption{Bayes at work}
\label{fig:bayes1}
\end{figure}

Figure~\ref{fig:bayes1} shows Eq.~\ref{eq:bayes} at work. Suppose $a$ is known to lie between 0 and 6, and
the prior distribution is taken as flat, as shown in the left hand plot.  A measurement of $a$ gives a result 
$4.4 \pm 1.0$~, as shown in the central plot. The product of the two gives (after normalization) the posterior, as shown in the right hand plot.

\subsubsection{Likelihood}

\label{sec:likelihood}
The likelihood---the number $P(Data|Theory)$---is now generalised to the function $L(a,x)$, where $x$ is the observed value of the data. Again, $x$ may be multidimensional, but in what follows it is not misleading to ignore that.
 
 This can be confusing.  For example, anticipating Section~\ref{sec:poisson}, the probability of getting $x$ counts from a Poisson process with mean $a$ is
 
\begin{equation}
 P(x,a)=e^{-a} {a^x \over x!}
 \label{eq:lone}
\quad.
\end{equation}
 
We also write
 \begin{equation}
 L(a,x)=e^{-a} {a^x \over x!}
  \label{eq:ltwo}
\quad.
\end{equation}

 What's the difference?  Technically there is none. These are identical joint functions of two variables ($x$ and $a$)
 to which we have just happened to have given different names.  Pragmatically we regard Eq.~\ref{eq:lone}
 as describing the probability of getting various different $x$ from some fixed $a$, whereas Eq.~\ref{eq:ltwo}
 describes the likelihood for various different $a$ from some given $x$. 
 But be careful with the term `likelihood'. If $P(x_1,a)>P(x_2,a)$ then $x_1$ is more probable (whatever you mean by that) than
 $x_2$. If $L(a_1,x)>L(a_2,x)$ it does not mean that $a_1$ is more likely (however you define that) than $a_2$.

 \subsubsection{Shortcomings of Bayesian probability}
 
 The big problem with Bayesian probability is that it is subjective.
Your $P_0(a)$ and my $P_0(a)$ may be different---so how can we compare results?
Science does, after all, take pride in being objective: it handles real facts, not opinions.
If you present a Bayesian result from your search for the $X$ particle this embodies
the actual experiment and your irrational prior prejudices.  I am interested in your experiment but not
in your irrational prior prejudices---I have my own---and it is unhelpful if you combine the two.

Bayesians sometimes ask about the right prior they should use. 
This is the wrong question.  The prior is what you believe, and only you know that.

There is an argument made for taking the prior as uniform. This is sometimes
called the 
`Principle of ignorance' and justified as being impartial. But this is misleading, even dishonest. 
If $P_0(a)$ is taken as constant, favouring no particular value, then it is not constant for $a^2$ or $\sqrt a$ or $\ln a$,
which are equally valid parameters.

It is true that with lots of data, $P_1(a)$ decouples from $P_0(a)$.
The final result depends only on the measurements.
But this is not the case  with little data---and that's the situation we're usually in---when doing statistics properly matters.

As an example, suppose you make a Gaussian measurement (anticipating slightly Section~\ref{sec:measurement}).
You consider a prior flat in $a$ and a prior flat in $\ln a$. This latter is quite sensible---it says you expect a 
result between 0.1 and 0.2 as being equally likely as a result between 1 and 2, or 10 and 20.
The posteriors are shown in Fig.~\ref{fig:differentpriors}.
For an `accurate' result of $3\pm 0.5$ the posteriors are very close. For an `intermediate' result
of $4.0 \pm 1.0$ there is an appreciable difference in the peak value and the shape. For a `poor'
measurement of $5.0 \pm 2.0$ the posteriors are {\em very} different.
 
\begin{figure}
\centerline{\includegraphics[width=8 cm, angle=270,trim={240 0 0 0},clip] {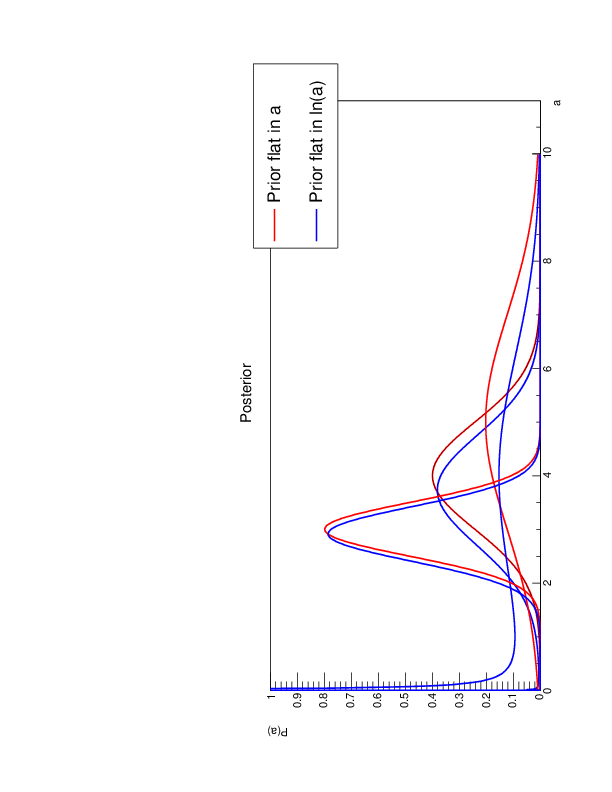}}

\caption{\label{fig:differentpriors} Posteriors for two different priors for the results $3.0 \pm 0.5$, $4.0 \pm1.0$ and $5.0 \pm 2.0$}
\end{figure}

 So you should never just quote results from a single prior. 
Try several forms of prior and examine the spread of results. If they are pretty much the same
you are vindicated. This is called
`robustness under choice of prior' and it is standard practice for statisticians. If they are different
then the data are telling you about the limitations of your results.

 \subsubsection {Jeffreys' prior}
\label{sec:Jeffreys}
 
Jeffreys~\cite{Jeffreys} suggested a technique now known as the Jeffreys' or {\em objective prior}: that
you should 
choose a prior flat in a transformed variable $a'$ for which the Fisher information, ${\cal I} =-\left< {\partial^2  L(x;a)
\over \partial a^2}\right> $ is constant. 
The Fisher information (which is important in maximum likelihood estimation, as described in Section~\ref{sec:ML})
is a measure of how much a measurement tells you about the parameter: a large ${\cal I}$ has a likelihood function with a sharp peak and will tell you (by some measure) a lot about $a$; a small ${\cal I}$ has a featureless likelihood function
which will not be useful. Jeffrey's principle is that the prior should not favour or disfavour particular values of the parameter.
It is equivalently---and more conveniently---used as taking a prior in the original $a$ which is proportional to
 $\sqrt{\cal I}$.

It has not been universally adopted for various reasons.  Some practitioners like to be able to include their own
prior belief into the analysis. It also makes the prior dependent on the experiment (in the form of the likelihood function). 
Thus if ATLAS and CMS searched for the same new $X$ particle they would use different priors for $P_0(M_X)$, 
which is (to some people) absurd.

So it is not universal---but when you are selecting a bunch of priors to test robustness---the Jefferys' prior 
is a strong contender for inclusion.

 \subsection{Summary}
 
 So mathematical probability has no meaning, and real probability is discredited. 
 That leaves the Frequentist and Bayesian definitions. Both are very much in use.
 
 They are sometimes presented as rivals, with adherents on either side (`frequentists versus Bayesians').
 This is needless drama. They are both tools that help us understand our results. Both have drawbacks. Sometimes it
 is clear which is the best tool for a particular job, sometimes it is not and one is free to choose either.
 It is said---probably accurately---that particle physicists feel happier with frequentist probability as they are used to
 large ensembles of similar but different events, whereas astrophysicists and cosmologists are more at home with Bayesian
 probability as they only have one universe to consider.
 
 What is important is not which version you prefer---these are not football teams---but that you know the limitations of each, that you use the best definition when there is a reason to do so, and, above all, that you are aware of which
 form you are using. 
 
 As a possibly heretical afterthought, perhaps  
classical probability still has a place?
Quantum Mechanics, after all, gives probabilities.
If $P_A$ is not `real'---either because it depends on an arbitrary ensemble,
or because is a subjective belief---then it looks like there is nothing `real' in the universe.

The state of a coin---or an electron spin---having probability $1 \over 2$ makes sense. There is a symmetry that dictates it.
The lifetime of a muon---i.e. probability per unit time that it will decay---seems to be a well-defined quantity, a property of the muon and independent of any ensemble, or any  Bayesian belief.

The probability a muon  will produce a signal in your muon detector seems like a `real  well-defined quantity', if you specify the 4 momentum and the state of the detector. Of course the inverse probability `What is the probability that a muon signal in my detector comes from a real muon, not background' is not intrinsically defined,  So
perhaps classical probability has a place in physics---but not in interpreting results.
However you should not mention this to a statistician or they will think you're crazy.

\section{Probability distributions and their properties} 

We have to make a simple distinction  between two sorts of data: \emph{integer} data and \emph{real-number} data\footnote{Other branches of science have to include a third, \emph{categorical} data, but we will ignore that.}.

The first covers results which are of their nature whole numbers: the numbers of kaons produced in 
a collision, or the number of entries  falling into some bin of a histogram.
Generically let's call such numbers $r$. They have probabilities $P(r)$ which are dimensionless.

The second covers results whose values are real (or floating-point) numbers.  There are lots of these:
energies, angles, invariant masses $\dots$
Generically let's call such numbers $x$, and they have probability density functions  $P(x)$
which have 
dimensions of $[x]^{-1}$, so   $\int_{x_1}^{x_2} P(x) dx$ or $P(x)\, dx$ are probabilities.

You will also sometimes meet the cumulative distribution  $C(x)=\int_{-\infty}^x P(x') \, dx'$.

 \subsection{Expectation values}

From $P(r)$ or $P(x)$ one can form the expectation value

\begin{equation}
\langle f \rangle =\sum_r f(r) P(r)  \qquad {\rm or} \qquad \langle f \rangle = \int f(x) P(x) \, dx 
\quad,
\end{equation}
where the sum or integral is taken as appropriate.
Some authors write this as
 $E(f)$, but I personally prefer the angle-bracket notation. You may think it looks too much like quantum mechanics,
 but in fact it's quantum mechanics which looks like statistics: an expression like $\langle \psi | \hat Q | \psi \rangle$
  is the average value of an operator $\hat Q$ in some state $\psi$, where `average value' has exactly the same 
  meaning and significance.

\subsubsection{Mean and standard deviation}

In particular  the {\em mean}, often written  $\mu$, is given by 

$ \langle r \rangle = \sum_r r P(r) \qquad  {\rm or } \qquad  \langle x \rangle =\int x P(x) \, dx \quad.$  

\noindent Similarly one can write 
higher {\em moments} 

 $\mu_k = \langle r^k \rangle = \sum_r r^k P(r)\qquad   {\rm or } \qquad  \langle x^k \rangle =\int x^k P(x) \, dx \quad,$  

\noindent and {\em central moments} 

$\mu'_k = \langle (r-\mu)^k \rangle = \sum_r (r-\mu)^k P(r) \qquad   {\rm or }  \qquad  \langle (x-\mu)^k \rangle =\int (x-\mu)^k P(x) \, dx \quad.$

\noindent The second central moment is known as the 
{\em variance}  

$\mu'_2=V= \sum_r (r-\mu)^2 P(r) = \langle r^2 \rangle - \langle r \rangle ^2$
  \qquad 
  or \qquad  $ \int (x-\mu)^2 P(x) \, dx =  \langle x^2 \rangle - \langle x \rangle ^2$

\noindent It is easy to show that $\langle (x-\mu)^2 \rangle =\langle x^2 \rangle -\mu^2$. The {\em standard deviation} is just the square root of the variance $\sigma=\sqrt{V}$.

Statisticians usually use variance, perhaps because formulae come out simpler. Physicists usually use standard deviation,
perhaps because it has the same dimensions as the variable being studied, and can be drawn as an error bar on a plot. 
  
  You may also meet {\em skew}, which  is $\gamma=\langle (x-\mu)^3 \rangle /\sigma^3$ and {\em kurtosis},   $h=\langle (x-\mu)^4  \rangle /\sigma^4 -3$.
  Definitions vary, so be careful. Skew is a dimensionless measure of the asymmetry of a distribution. Kurtosis is
  (thanks to that rather arbitrary looking 3 in the definition) 
  zero for a Gaussian distribution (see Section~\ref{sec:measurement}): positive kurtosis indicates a
  narrow core with a wide tail, negative kurtosis indicates the tails are reduced.

  \subsubsection{Covariance and correlation}

If your data are 
2-dimensional pairs $(x,y)$,
then besides forming  $\langle x \rangle, \langle y \rangle, \sigma_x$ etc., you can also form the 
 {\em Covariance}
 
  ${\rm Cov}(x,y)=\langle (x-\mu_x)(y-\mu_y) \rangle = \langle xy \rangle - \langle x \rangle \langle y \rangle \quad.$

 \begin{figure}
\centerline{
\includegraphics[width=7 cm, angle=270,trim={0 0 0 0 },clip]{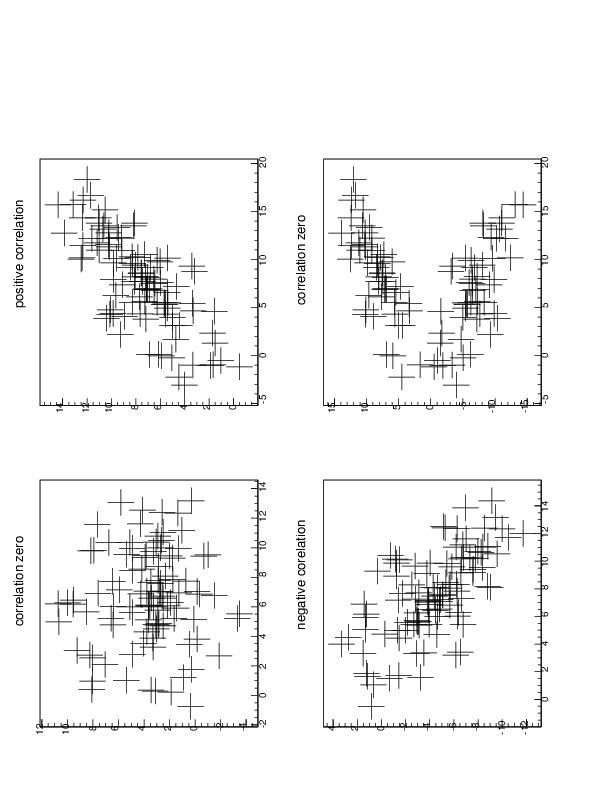}
}

\caption{\label{fig:covariance} Examples of two dimensional distributions. The top right  has positive covariance (and correlation), the bottom
left negative. In the top left the covariance is zero and $x$ and $y$ are independent; in the bottom right
the covariance is also zero, but they are not independent.}
\end{figure}

Examples are shown in Fig.~\ref{fig:covariance}. If there is a tendency for positive fluctuations in $x$ to be associated with positive fluctuations in $y$ (and therefore negative with negative) then
the product $(x_i-\overline x)(y_i-\overline y)$ tends to be  positive and the covariance is greater than 0. A negative covariance, as in the 3rd plot, happens if a positive fluctuation in one variable is associated with a negative fluctuation in the other.
If the variables are independent then a positive variation in $x$ is equally likely to be associated with a positive or a negative variation in $y$ and the covariance is zero, as in the first plot. However the converse is not always the case, there can be two-dimensional distributions where the covariance is zero, but the two variables are not independent, as is shown in the fourth plot.

 Covariance is useful, but it has dimensions. Often one uses the 
 {\em correlation}, which is just
 \begin{equation}
 \rho={{\rm Cov}(x,y)\over  \sigma_x \sigma_y}
\quad.
\end{equation}

It is easy to show that
$\rho$ lies between 1 (complete correlation) and -1 (complete anticorrelation). 
$\rho=0$ if $x$ and $y$ are independent.
 
 If there are more than two variables---the alphabet runs out so let's call them 
$(x_1,x_2,x_3\dots x_n)$---
then these generalise to the 
{\em covariance matrix}

 ${\bf V}_{ij}=\langle x_i x_j \rangle - \langle x_i \rangle \langle x_j \rangle$ 

\noindent and the {\em 
correlation matrix}

 ${\bf \rho}_{ij}={{\bf V}_{ij} \over \sigma_i \sigma_j} \quad.$

\noindent The diagonal of $\bf V$ is $\sigma_i^2$.  
The
diagonal of $\bf \rho$ is 1.

\subsection{Binomial, Poisson and Gaussian}
 
 We now move from considering the general properties of distributions to considering three specific ones.
 These are the ones you will most commonly meet for the distribution of the original data
 (as opposed to quantities constructed from it). Actually the first, the binomial, is not nearly as common as the second, the Poisson; and the third, the Gaussian, is overwhelmingly more common. However it is 
 useful to consider all three as concepts are built up from the simplest to the more sophisticated.
 
\subsubsection  {The binomial distribution  }

The binomial distribution is easy to understand as it basically describes the familiar\footnote{Except, as it happens, in Vietnam, where coins have been completely replaced by banknotes.} tossing of coins.    
It describes the number $r$  of successes in $N$ trials, each with probability $p$ of success.
$r$ is discrete so the process is described by a probability distribution

\begin{equation}
P(r;p,N)={N! \over r! (N-r)!} p^r q^{N-r} 
\quad,
\end{equation}
where $ q \equiv 1-p$.

Some examples are shown in Fig.~\ref{fig:binom}.
 
\begin{figure}[ht]
\centerline{
\includegraphics[width=7 cm, angle=270, trim={200 0 0 0} , clip ] {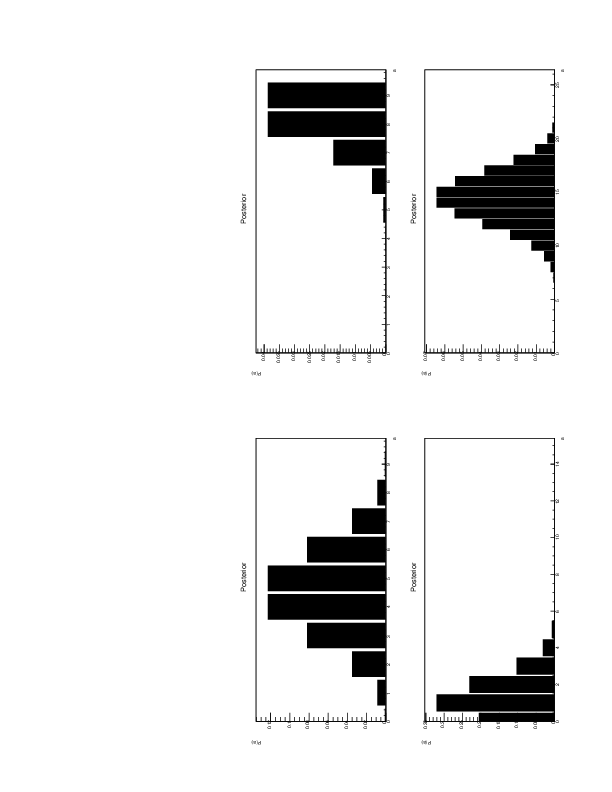}
}
\caption{\label{fig:binom} Some examples of the binomial distribution, for
(1) $N=10,p=0.6$,
(2) $N=10, p=0.9$,
(3) $N=15, p=0.1$,
and 
(4) $N=25,p=0.6$.
}
\end{figure}

The distribution has 
mean $\mu=Np$, variance $V=Npq$,  and standard deviation $\sigma=\sqrt{Npq}$.
 
 \subsubsection {The Poisson distribution}

The Poisson distribution also describes the probability of some discrete number $r$,
but rather than a fixed number of `trials' it considers a  random rate $\lambda$: 
\label{sec:poisson}
\begin{equation}
P(r;\lambda)=e^{-\lambda}{\lambda^r \ \over r!}
\quad.
\end{equation}

It is linked to the binomial---the Poisson is the 
limit of the binomial---as $N\to \infty$, $p \to 0$ with $np=\lambda=constant$. Figure~\ref{fig:poisson} shows various examples. It has mean $\mu=\lambda$, variance $V=\lambda$, and standard deviation $\sigma=\sqrt{\lambda}=\sqrt{\mu}$.
 
 \begin{figure}[ht]
\centerline{
\includegraphics[width=7 cm, angle=270, trim={200 0 0 0} , clip ] {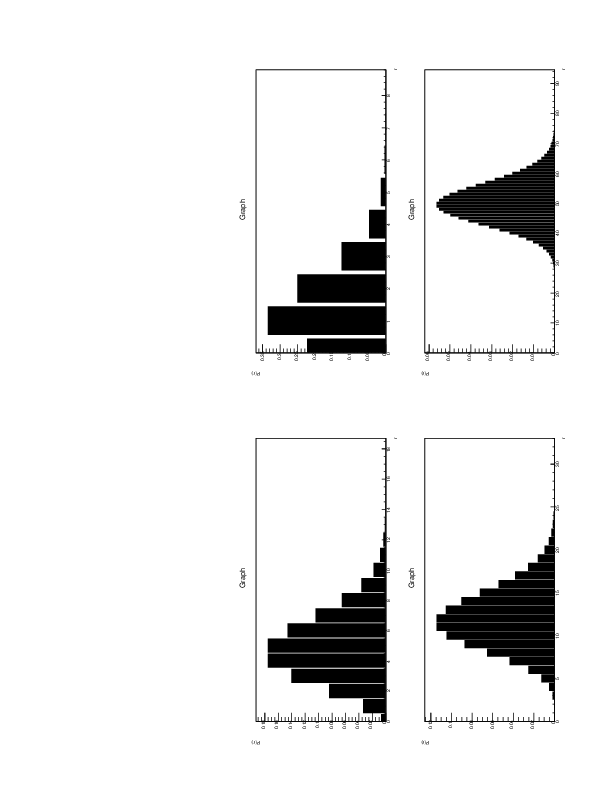}}
\caption{\label{fig:poisson} Poisson distributions for
(1) $\lambda=5$,
(2) $\lambda=1.5$,
(3) $\lambda=12$ and
(4) $\lambda=50$
}
\end{figure}

The clicks of a Geiger counter are the standard illustration of a Poisson process.
You will meet it  a lot as it applies to event counts---on their own or in histogram bins.
 
To help you think about the Poisson, here is a simple question (which
describes a situation 
I have seen in practice, more than once, from people who ought to know better).
 
   \
 
 \hrule
 
 \

You need to know the efficiency of your PID  system for positrons.

You find 1000 data events where 2 tracks have a combined mass of 3.1~GeV ($J/\psi)$ and the negative track is 
identified as an $e^-$ (`Tag-and-probe' technique).

In 900 events the $e^+$ is also identified. In 100 events it is not. The efficiency is 90\%.

What about the error?

Colleague A says  $\sqrt{900}=30$ so efficiency is $90.0 \pm 3.0 $\%,

colleague B says $\sqrt{100}=10$ so efficiency is $90.0  \pm 1.0 $\%.

Which is right?

 \
 
 \hrule
 
 \
 
 Please think about this before turning the page...
 
 \vfill\eject
 
 {Neither---both are wrong}. This is binomial not Poisson: $p=0.9, N=1000$.

\noindent The error is $\sqrt{Npq}=\sqrt{1000 \times 0.9 \times 0.1}$ (or $\sqrt{1000 \times 0.1 \times 0.9}$) =$\sqrt{90} = 9.49$ so the efficiency is $90.0 \pm 0.9$ \%.  

  \subsubsection{The Gaussian distribution}
 
 This is by far the most important statistical distribution.
 \label{sec:measurement}
 The probability density function (PDF) for a variable $x$ is given by the formula
 \begin{equation} 
P(x;\mu,\sigma)={1 \over \sigma \sqrt{2 \pi}} e^{-{(x-\mu)^2 \over 2 \sigma^2}}
\quad.
\end{equation}

Pictorially this is shown in Fig.~\ref{fig:gauss}.

\begin{figure}[h]

\centerline{ \includegraphics[width=8 cm, angle=270, trim={200 0 0 0} , clip ] {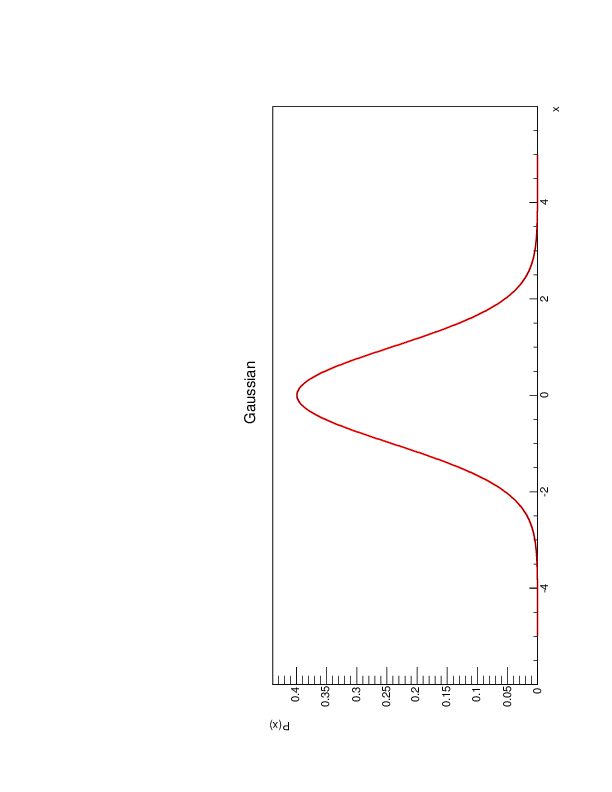}}
\caption{\label{fig:gauss}The Gaussian distribution}
\end{figure}
 
 This is sometimes called the `bell curve', though in fact a real bell does not have flared edges like that.
 There is (in contrast to the Poisson and binomial)  
 only one 
 Gaussian curve, as $\mu$ and $\sigma$ are just location and scale parameters.

The mean is $\mu$ and the standard deviation is  $\sigma$. The 
 Skew is zero, as it is symmetric, and the kurtosis is zero by construction.
 
 In statistics, and most disciplines, this is known as the {\em normal distribution}. Only in physics is it known as `The Gaussian'---perhaps because the word `normal' already has so many meanings. 
 
 The reason for the importance of the Gaussian is the {\em central limit theorem} (CLT) that states:
if the variable $X$ is the sum of $N$  variables $x_1,x_2\dots x_N$ then:

\begin{enumerate}
\item Means add:  $ \langle X \rangle = \langle x_1 \rangle + \langle x_2 \rangle + \dots \langle x_N \rangle$, 
\item Variances add: $V_X=V_1+V_2 +\dots V_N$,
\item  If the variables $x_i$ are independent and identically distributed (i.i.d.) then $P(X)$ tends to a Gaussian for large $N$.
\end{enumerate}

(1) is obvious, (2) is pretty obvious, and means that standard deviations add in quadrature, and that the standard deviation of an average falls like $1\over \sqrt N$, (3) applies whatever the form of the original  $P(x)$.

Before proving this, it is helpful to see a demonstration to convince yourself that the implausible assertion in (3)
actually does happen.
Take a uniform distribution from 0 to 1, as shown in the top left subplot of Fig.~\ref{fig:CLT}. It is flat. Add two such numbers and the distribution is triangular, between 0 and 2, as shown in the top right.

\begin{figure}[h]
\centerline{\includegraphics[width=8 cm, angle=270, trim={200 0 0 0} , clip ] {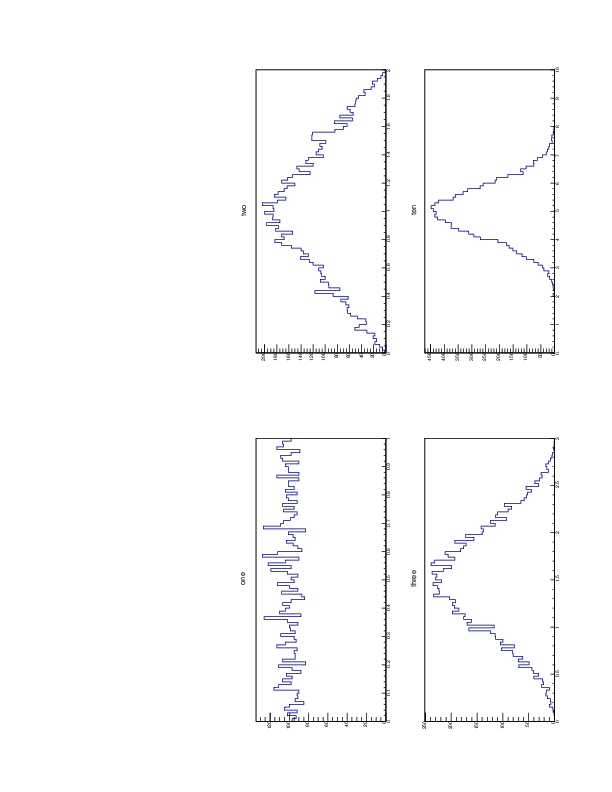}}
\caption{\label{fig:CLT}Demonstration of the central limit theorem}
\end{figure}
With 3 numbers, at the bottom left, it gets curved. With 10 numbers, at the bottom right, it looks pretty Gaussian. The proof follows.

\begin{proof}
 
First, introduce the characteristic function  $\langle e^{i k x} \rangle = \int e^{i k x } P(x) \, dx = \tilde P(k)$.

This can usefully be thought of as an expectation value and as a Fourier transform, FT.

Expand the exponential as a series

$\langle e^{i k x} \rangle = \langle 1+ikx+{(ikx)^2 \over 2!}+{(ikx)^3 \over 3!}\dots \rangle = 1 + ik \langle x \rangle +(ik)^2{\langle x^2 \rangle \over 2!} + (ik^3) {\langle x^3 \rangle \over 3!} \dots$. 
Take the logarithm and use the expansion $\ln(1+z)=z-{z^2 \over 2 } + {z^3 \over 3} \dots$
This gives a power series in $(ik)$, where the coefficient ${\kappa_r \over r!}$ of $(ik)^r$ is made up of expectation values of $x$ of total power $r$

$\kappa_1= \langle x \rangle, \kappa_2= \langle x^2 \rangle - \langle x \rangle ^2 =, \kappa_3= \langle x^3 \rangle -3 \langle x^2 \rangle \langle x \rangle +2 \langle x \rangle^3 \dots$ 
 
These are called the semi-invariant cumulants of Thi\`ele . Under a change of scale $\alpha$, $\kappa_r \to \alpha^r \kappa_r$. Under a change in location only $\kappa_1$ changes.
 
If $X$ is the sum of i.i.d. random variables, $x_1+x_2+x_3...$, then $P(X)$ is the convolution of $P(x)$ with itself $N$ times.
 
The FT of a convolution is the  product of the individual FTs,

the logarithm of a product is the sum of the logarithms,
  
so $P(X)$ has cumulants $K_r=N \kappa_r$.

 To make graphs commensurate, you need to scale the $X$ axis by the
standard deviation, which grows like $\sqrt{N}$.  The cumulants of the scaled graph are $K'_r = N^{1-r/2} \kappa_r$. 

As $N \to \infty$, these vanish for $r>2$, leaving a quadratic.
  
 If the log is a quadratic, the exponential is a Gaussian. So $\tilde P(X)$ is Gaussian.
 
 And finally, the inverse FT of a Gaussian is also a Gaussian.
  \end{proof}    
 
Even if the distributions are not identical, the CLT tends to apply, unless one (or two) dominates.
Most `errors' fit this, being compounded of many different sources.

  \section{Hypothesis testing}
   
 `Hypothesis testing' is another piece of statistical technical jargon. 
It just means `making choices'---in a logical way---on the basis of statistical information. 

\begin{itemize}
\item
Is some track a pion or a kaon?
\item Is this event signal or background?
\item Is the detector performance degrading with time?
\item Do the data agree with the Standard Model prediction or not?
\end{itemize}

To establish some terms: you have a {\it hypothesis} (the track is a pion, the event is signal,
the detector is stable, the Standard Model is fine $\dots$). and an alternative hypothesis (kaon, background, changing, new physics needed $\dots$)  Your hypothesis is usually {\it simple} i.e. completely specified, 
but the alternative is often {\it composite} containing a parameter (for example, the detector decay rate) which may have any non-zero value.

 \subsection{Type I and type II errors}

As an example, let's use the signal/background decision. Do you accept or reject the event (perhaps in the trigger, perhaps in your offline analysis)? To make things easy we consider the case where both hypotheses are simple, i.e. completely defined.

Suppose you measure some parameter $x$ which is related to what you are trying to measure.
It may well be the output from a neural network or other machine learning (ML) systems. 
The expected distributions for $x$ under the hypothesis and the alternative, $S$ and $B$ respectively, are shown in Fig.~\ref{fig:hyp}.  

\begin{figure}[h]
\centerline{\includegraphics[width=6 cm, angle=270, trim={200 0 0 0} , clip ] {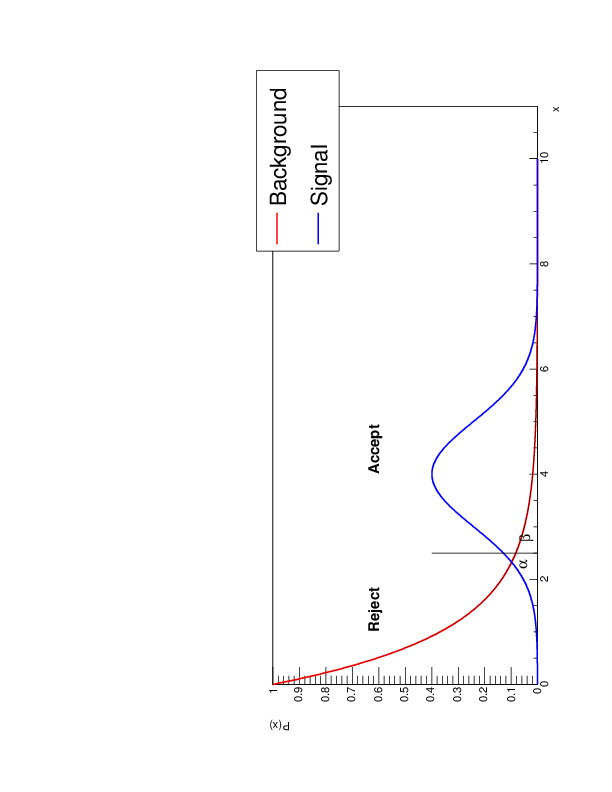}}
\caption{\label{fig:hyp} Hypothesis testing example}
\end{figure}
 
You impose a cut as shown---you have to put one somewhere---accepting events above $x=x_{cut}$ and rejecting those below.

This means losing 
a
fraction $\alpha$ of signal.   This is called a {\em type I error} and  $\alpha$ is known as  the {\em significance}.

You admit a  fraction $\beta$ of background. This is called a {\em type II error} and  $1-\beta$ is the power.
 
 You would like to know the best place to put the cut. This graph cannot tell you!   
 The strategy for the cut depends on three things---hypothesis testing only covers one of them.
 
The second  is the 
 prior signal to noise ratio.
These  plots are normalized to 1. The red curve is (probably) MUCH bigger.
A value of $\beta$ of, say, 0.01 looks nice and small---only one in a hundred background events get through.
But if your background is 10,000 times bigger than your signal (and it often is) you are still swamped.
 
 The third is the cost of making mistakes, which will be different for the two types of error.
 You have a trade-off between efficiency and purity: what are they worth?
 In a typical analysis, a type II error is more serious than a type I: losing a signal event is regrettable, but it happens. 
 Including background events in your selected pure sample can give a very misleading result. 
 By contrast,  
 in medical decisions, type I errors are  much worse than type II.  Telling healthy patients they are sick leads to worry and perhaps further tests, but telling sick patients they are healthy means they don't get the treatment they need.


\subsection  {The Neymann-Pearson lemma}

In Fig.~\ref{fig:hyp} the strategy is plain---you choose $x_{cut}$ and evaluate $\alpha$ and $\beta$.
But
suppose the $S$ and $B$ curves are more complicated, as in Fig.~\ref{fig:hyp1}? Or that $x$ is multidimensional?

\begin{figure}[h]
\centerline{
\includegraphics[width=6 cm, angle=270, trim={250 0 0 0} , clip ] {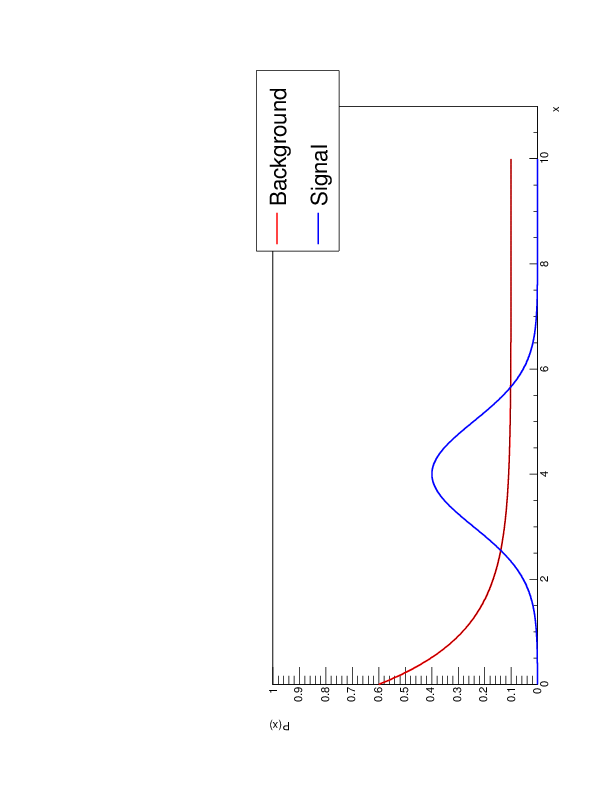}}
\caption{\label{fig:hyp1} A more complicated case for hypothesis testing}
\end{figure}

Neymann and Pearson  say: your acceptance region just includes regions of greatest $S(x) \over B(x)$  (the ratio of likelihoods).
For a given $\alpha$, this gives the smallest $\beta$   (`Most powerful at a given significance')

The proof is simple: having done this,  if you then move a small region from `accept' to `reject' it has to be replaced by an equivalent region, to balance $\alpha$,  which (by construction) 
brings more background, increasing $\beta$.

However complicated, such a problem reduces to a single monotonic variable $S \over B$, and you cut on that. 

 \subsection{Efficiency, purity, and ROC plots}

\begin{figure}[h]
\centerline{
\includegraphics[width=5 cm, angle=270, trim={250 0 0 0} , clip ] {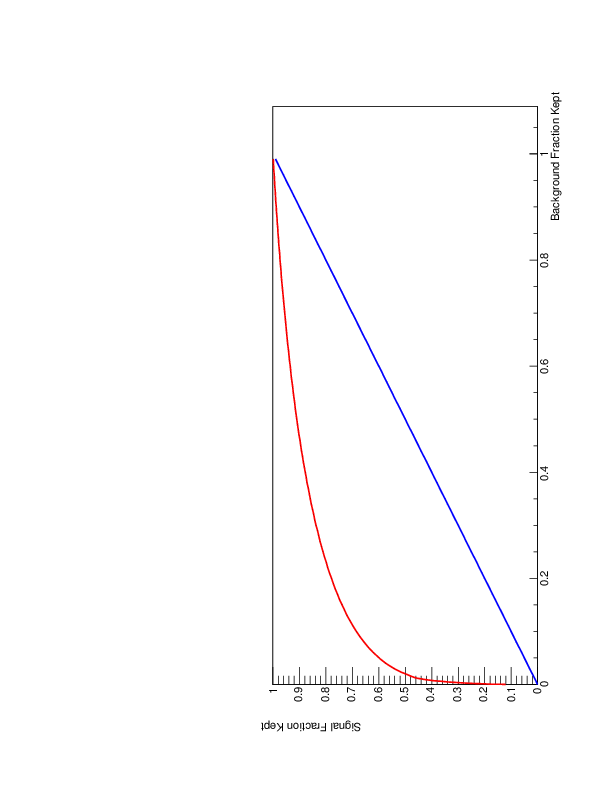}
}
\caption{\label{fig:ROC} ROC curves}

\end{figure}

ROC plots are often used to show the efficacy of different selection variables.
You scan over the cut value (in $x$, for Fig.~\ref{fig:hyp} or in $S/B$ for a case like Fig.~\ref{fig:hyp1}
and plot the  fraction of background accepted ($\beta$)  against fraction of signal retained ($1-\alpha$),
as shown in Fig.~\ref{fig:ROC}. 

For a very loose cut all data is accepted, corresponding to a point at the top right. As the cut is tightened both signal and background fractions fall, so the point moves  to the left and down, though hopefully the background loss is greater than the signal loss, so it moves more to the left than it does downwards.  As the cut is increased the line moves towards the bottom left, the limit of a very tight cut where all data is rejected.

A diagonal line corresponds to no discrimination---the  $S$ and $B$ curves are identical.
The further the actual line bulges away from that diagonal, the better.  

Where you should put your cut depends, as pointed out earlier, also on the prior signal/background ratio and the relative costs of errors. The ROC plots do not tell you that, but they can be useful in comparing the performance of different
discriminators.

The name `ROC' stands for  
`receiver operating characteristic', for reasons that are lost in history. Actually it is good to use this meaningless acronym, otherwise they get called `efficiency-purity plots' even though they definitely do not show the purity (they cannot, as that  depends on the overall signal/background ratio).   Be careful, as the phrases
 `background efficiency', `contamination', and `purity' are used ambiguously in the literature.

  \subsection{The null hypothesis}
 
An analysis is often (but not always) investigating whether an effect is present, motivated by
the hope that the results will show that it is:

\begin{itemize}
\item Eating broccoli makes you smart.
\item Facebook advertising increases sales.
\item A new drug increases patient survival rates.
\item The data show Beyond-the-Standard-Model physics.
\end{itemize}

To reach such a conclusion you have to use your best efforts to try, and to fail, to prove the opposite: the {\em Null Hypothesis} $H_0$.

\begin{itemize}
\item Broccoli lovers have the same or small IQ than broccoli loathers.
\item Sales are independent of the Facebook advertising budget.
\item The survival rates for the new treatment is the same.
\item The Standard Model (functions or Monte-Carlo) describe the data.
\end{itemize}

If the null hypothesis is not tenable, you've proved---or at least, supported---your point.   

The reason for calling $\alpha$  the `significance' is now clear. It is the probability that the null hypothesis will be wrongly rejected, and you'll claim an effect where there isn't any.

There is a minefield of difficulties. Correlation is not causation. If broccoli eaters are more intelligent, 
perhaps that's because it's intelligent to eat green vegetables, not that vegetables make you intelligent. 
One has to consider that if similar experiments are done, self-censorship will influence which results get published. 
This is further discussed in  Section~\ref{sec:discovery}.

This account is perhaps unconventional in introducing the null hypothesis at such a late stage. Most treatments
bring it in right at the start of the description of hypothesis testing, because they assume that all decisions are of this type.

\def \xbar {\overline x}
\def \xsqbar {\overline {x^2}}

\section{Estimation}

 What statisticians call `estimation',
physicists would generally call  `measurement'.

 Suppose you
 know the probability (density) function  $P(x;a)$ 
and you 
take  a set of data $\{x_i\}$.   What is the best value for $a$? (Sometimes one wants to estimate a property (e.g. the mean) rather than a parameter, but 
this is relatively uncommon, and the methodology is the same.) 
 
$x_i$ may be single values, or pairs, or higher-dimensional.
The unknown
$a$ may be a single parameter or several. If it has more than one component, these are sometimes split into `parameters of interest' and `nuisance parameters'.

 The {\em estimator}  is defined very broadly:
an estimator $\hat a(x_1\dots x_N)$  is a function of the data that gives a value for the parameter $a$. There is no `correct' estimator, but some are better than others. A perfect estimator would be:

\begin{itemize}
\item
Consistent.   $\hat a(x_1 \dots x_N) \to a$ as $ N \to \infty $,
\item
Unbiased:  $\langle \hat a \rangle = a $,
\item
 Efficient:  $\langle (\hat a - a)^2 \rangle$ is as small as possible,
\item 
  Invariant:  $\hat f(a) = f(\hat a)$.
\end{itemize}

No estimator is perfect---these 4 goals are incompatible. In particular the  second and the fourth; if
an estimator $\hat a$ is unbiased for $a$ then
$\sqrt{\hat a}$ is not an unbiased estimator of $\sqrt a$.

\subsection{Bias}

Suppose we estimate the mean by taking the obvious\footnote{Note the difference between $\langle x \rangle$  which is an average over a PDF and $\overline x$ 
which denotes the average over a particular sample: both are called `the mean $x$'.} $\hat \mu = \xbar$

$\left< \hat \mu \right> = \left< {1 \over N } \sum x_i \right> = {1 \over N } \sum \mu = \mu$. 

So there is no bias. This expectation value of this estimator of $\mu$ is just $\mu$ itself. By contrast suppose
we estimate the variance by the apparently obvious
 $\hat V = \xsqbar-\xbar^2$.

Then $\left<  \hat V \right> = \left< \xsqbar \right> - \left< \xbar^2 \right>$.

The first term is just $\left< x^2 \right>$. To make sense of the second term, note that $\left< x \right> = \left< \xbar \right>$ and add and subtract $\left< x \right>^2$ to get

$\left<  \hat V \right> = \left<  x^2 \right> - \left< x \right>^2 -  (\left< \xbar^2 \right>- \left< \xbar \right>^2)$

$\left<  \hat V \right> =V(x)-V(\xbar)=V-{V \over N}={N-1 \over N} V$.

So the estimator is biased! $\hat V$ will, on average, give too small a value.

This bias, like any known bias, can be corrected for. 
  Using $\hat V = {N \over N-1} (\xsqbar-\xbar^2)$ corrects the bias. The familiar estimator for
  the standard deviation follows:
  $\hat \sigma=\sqrt{\sum_i (x_i-\xbar)^2 \over N-1}$. 
  
  (Of course this gives a biased estimate of $\sigma$.  But $V$ is generally more important in this context.)

\subsection {Efficiency}
\label{sec:ML}
 
 Somewhat surprisingly, there is a limit to the efficiency of an estimator: the
 {\em  minimum variance bound} (MVB),
also known as the  {\em Cramér-Rao bound}.

For any unbiased estimator $\hat a(x)$, the variance is bounded 

\begin{equation}V(\hat a)\geq
-{1 \over \left< {d^2 \ln L \over da^2}\right>}
={1 \over \left<
\left({d \ln L \over da }\right) ^2
\right>}
\quad.
\end{equation}

$L$ is the likelihood (as introduced in Section~\ref{sec:likelihood}) of a sample of independent measurements, i.e. the
probability for the whole data sample for a particular value of $a$.
It is just the product of the individual probabilities:

 $L(a;x_1,x_2,...x_N)=P(x_1;a)P(x_2;a)...P(x_N;a)$.

We will write $L(a;x_1,x_2,...x_N)$ as $L(a;x)$ for simplicity.

\begin{proof}{Proof of the MVB}

Unitarity requires $\int P(x;a)\, dx = \int L(a;x) \, dx =1$

Differentiate wrt $a$: \qquad  \begin{equation}
0=\int {dL \over da} \, dx = \int L {d \ln L \over da} \, dx = \left<  {d \ln L \over da} \right>
\label{eq:one}
 \end{equation}

If $\hat a$ is unbiased 
$\left< \hat a\right> = \int \hat a(x) P(x;a) \, dx = \int \hat a(x) L(a;x) \, dx =a$

Differentiate wrt $a$: \qquad  $1=\int \hat a(x) {dL \over da} \, dx = \int \hat a L {d \ln L \over da} \, dx  $
 
 Subtract    Eq.~\ref{eq:one} multiplied by $a$, and     get $\int (\hat a - a){d \ln L \over da} L dx =1$
 
 Invoke the Schwarz inequality $\int u^2 \, dx \int v^2 \, dx \geq \left( \int u v \, dx \right)^2 $ with $u\equiv (\hat a - a) \sqrt L, v\equiv {d \ln L \over da} \sqrt L$
 
 Hence $\int (\hat a - a)^2 L \, dx  \int \left( {d \ln L \over da}\right)^2 L \, dx \geq 1$

 \begin{equation} 
 \left< (\hat a - a)^2 \right> \geq 1/\left<\left( {d ln L \over da }\right)^2 \right>
 \end{equation}
 
  \end{proof}

Differentiating Eq.~\ref{eq:one} again gives

${ d \over da} \int L { d \ln L \over da} \, dx = \int {d L \over da} \, {d \ln L \over da} \, dx + \int L {d^2 \ln A \over da^2} \, dx
=
\left< \left( {d \ln L \over da} \right)^2\right>+\left<{d^2 \ln L \over da^2}\right>=0$,

hence 
$\left< \left( {d \ln L \over da} \right)^2\right>= - \left<{d^2 \ln L \over da^2}\right>$.

This is the  {\em Fisher information} referred to in Section~\ref{sec:Jeffreys}. Note how it is intrinsically positive.

\subsection{Maximum likelihood estimation}

 The {\em maximum likelihood} (ML) estimator just does what it says: $a$ is adjusted to maximise the
 likelihood of the sample
(for practical reasons one actually maximises the log likelihood, which is a sum rather than a product).

\begin{equation}
{\rm Maximise } \ln L = \sum_i \ln {P(x_i;a)}
\label{eq:logL}
\quad,
\end{equation}

\begin{equation}
\left. {d \ln L \over d a } \right|_{\hat a}=0
\label{eq:dlogL}
\quad.
\end{equation}

The 
ML estimator is very commonly used. It is not only simple and intuitive, it has lots of nice properties.

\begin{itemize}
\item
It is consistent.

\item
It is biased, but bias falls like $1/N$.

\item
It is efficient for the large $N$.

\item
It is invariant---doesn't matter if you reparametrize  $a$. 
\end{itemize}

A particular maximisation problem may be solved in 3 ways, depending on the complexity

\begin{enumerate}
\item Solve Eq.~\ref{eq:dlogL} algebraically,
\item Solve Eq.~\ref{eq:dlogL} numerically, and
\item Solve Eq.~\ref{eq:logL} numerically.
\end{enumerate}

\subsection{Least squares}

 {\em Least squares estimation} follows from maximum likelihood estimation.
If you have 
Gaussian measurements of $y$ taken at various $x$ values, with measurement error $\sigma$, and a prediction $y=f(x;a)$
then the Gaussian probability

\centerline{$P(y;x,a)={1 \over \sigma \sqrt{2 \pi}} e^{-(y-f(x,a))^2/2 \sigma^2}$}

gives the log likelihood

\centerline{$\ln L = - \sum { \left(y_i - f(x_i;a)\right)^2  \over 2 \sigma_i^2} + {\rm constants}$.}

To maximise $\ln L$, you minimise  $\chi^2 =  \sum { \left(y_i - f(x_i;a)\right)^2  \over  \sigma_i^2} $, hence the name `least squares'.

Differentiating gives the {\em normal equations}:
$\sum { \left(y_i - f(x_i;a)\right)  \over  \sigma_i^2}f'(x_i;a) =0$.

If $f(x;a)$ is linear in $a$ then these can be solved exactly. Otherwise an iterative method has to be used.

\subsection{Straight line fits}

As a particular instance of least squares estimation, suppose the function is $y=mx+c$, and assume all $\sigma_i$  are the same (the extension to the general case is straightforward).
The normal equations are then $\sum   (y_i - m x_i -c) x_i = 0$ and $\sum  (y_i-m x_i - c ) =0$\ , for which the solution, shown in Fig.~\ref{fig:slfit}, is

\noindent $m={\overline{xy} - \overline x \ , \overline y \over \xsqbar - \xbar^2}$\ , $c=\overline y - m \xbar$ \ .
 
 \begin{figure}[h]
 \begin{center}
\includegraphics[width=5 cm, angle=270, trim={0 0 0 0},clip ] {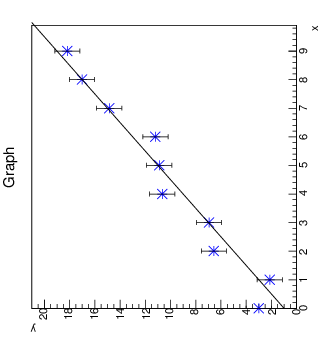}
 \caption{\label{fig:slfit} A straight line fit}
 \end{center}
 \end{figure}
 
 Statisticians call this {\em  regression}. Actually there is a subtle difference, as shown in Fig.~\ref{fig:regression}.

 \begin{figure}[h]
 \begin{center}
\includegraphics[width=5 cm, angle=270, trim={260 0 0 0},clip ] {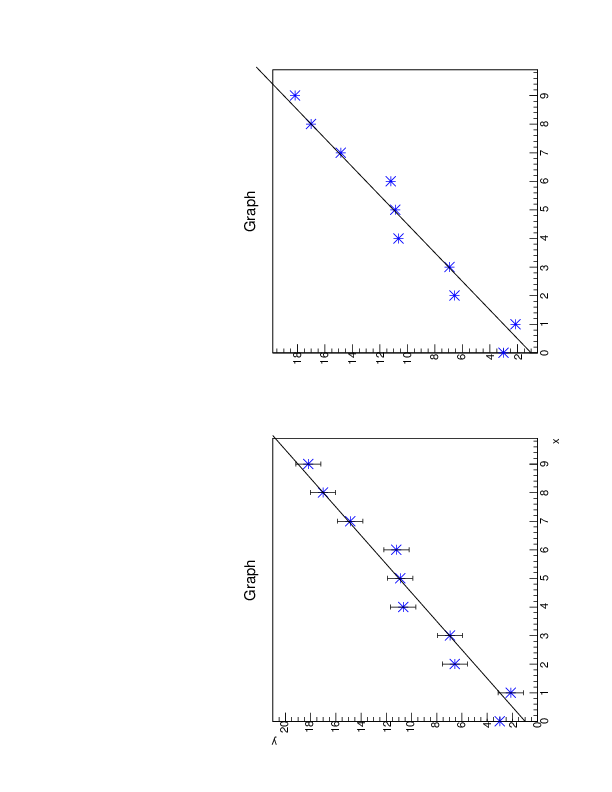}
\caption{\label{fig:regression} A straight line fit (left) and linear regression (right)}
\end{center}
 \end{figure}
 
 The straight line fit considers well-defined $x$ values and $y$ values with measurement errors---if it were not for those
 errors then presumably the values would line up perfectly, with no scatter. The scatter in regression is not caused by measurement errors, but by the fact that the variables are linked only loosely.  
  
 The history of regression started with Galton, who measured the heights of fathers and their (adult) sons.
 Tall parents tend to have tall children so there is a correlation. Because the height of a son depends
 not just on his paternal genes but on many factors (maternal genes, diet, childhood illnesses $\dots$), the points
 do not line up exactly---and using a high accuracy laser interferometer to do the measurements, rather than a simple
 ruler, would not change anything. 
 
 Galton, incidentally, used this to show that although    
tall fathers tend to have tall sons, they are not that tall. An outstandingly tall father will have (on average) quite tall children, and only tallish grandchildren. He called this 
 `Regression towards mediocrity', hence the name.
 
 It is also true that  tall sons tend to have tall fathers---but not that tall---and only tallish grandfathers. Regress works in both directions!

Thus for regression there is always an ambiguity as  to whether to plot $x$ against $y$ or $y$ against $x$.
For a straight line fit as we usually meet them this does not arise: one variable is precisely specified and we call that one $x$, and the one with measurement errors is  $y$.

 \subsection{Fitting histograms}

When fitting a histogram the error is given by Poisson statistics for the number of events in each bin.

 \begin{figure}
 \centerline{\includegraphics[width=6 cm, angle=270, trim={260 0 0 0},clip ] {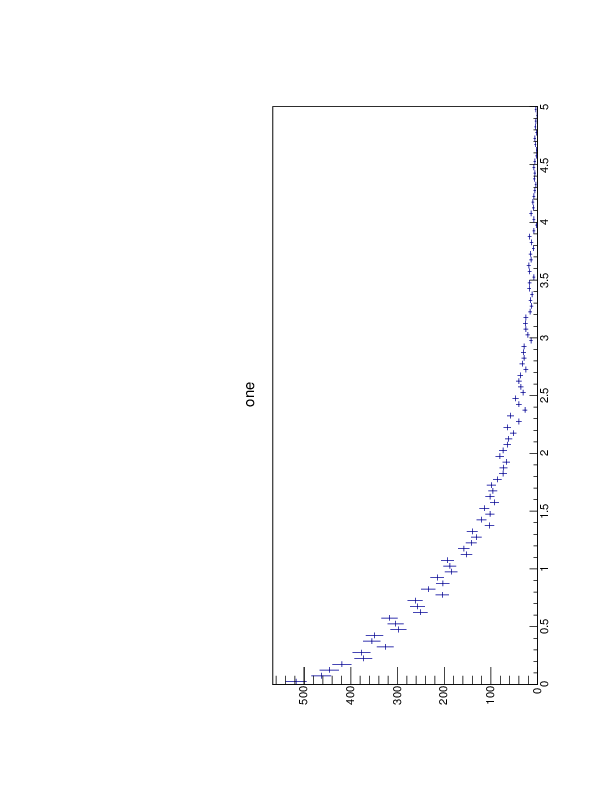}}
\caption{\label{fig:fithist} Fitting a histogram}
 \end{figure}

There are 
4 methods of approaching this problem---in order of increasing accuracy and decreasing speed.   It is assumed that the bin width $W$  is narrow, so that $f(x_i,a)=\int_{x_i}^{x_i+W} P(x,a)\, dx$ can be approximated by 
 $f_i(x_i;a)=P(x_i;a) \times W$. $W$ is almost always the same for all bins,
but the rare cases of variable bin width can easily be included.
\begin{enumerate}
\item Minimise $\chi^2 = \sum_i {(n_i-f_i)^2 \over n_i}$. This is the simplest but clearly breaks if  $n_i=0$.
\item Minimise $\chi^2 = \sum_i {(n_i-f_i)^2\over f_i}$ . Minimising the Pearson $\chi^2$ (which {\em is}
valid here) avoids the division-by-zero problem. It assumes that the Poisson distribution can be approximated by a Gaussian.
\item Maximise $\ln L = \sum \ln(e^{-f_i} f_i^{n_i} / n_i!) \sim \sum n_i \ln f_i - f_i$. This, known as {\em binned maximum likelihood},  remedies that assumption.
\item Ignore bins and maximise  the total likelihood. Sums run over $N_{events}$ not $N_{bins}$. So if you have large data samples this is much slower. You have to use it for sparse data, but of course in such cases the sample is small and the
time penalty is irrelevant.
\end{enumerate}

Which method to use is something you have to decide on a case by case basis. 
If you have bins with zero entries then the first method is ruled out
(and removing such bins from the fit introduces bias so this should not be done).
Otherwise, in my experience, the improvement in adopting a more complicated method tends to be small.

 \section{Errors}
 
 Estimation gives you a value for the parameter(s) that we have called $a$. But you also---presumably---want to 
 know something about the uncertainty on that estimate. The maximum likelihood method provides this.
 
 \subsection{Errors from likelihood}

For large $N$, the $\ln L(a,x)$ curve is a parabola, as shown in Fig.~\ref{fig:ML}.

\begin{figure} [h]
\centerline{
\includegraphics[width=6.5 cm, angle=270, trim={240 0 0 0},clip ] {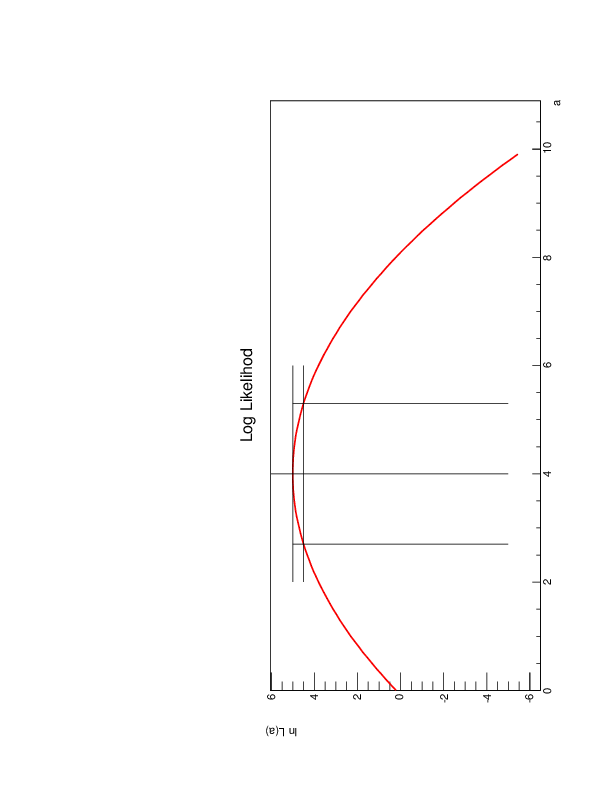}}
\caption{\label{fig:ML} Reading off the error from a Maximum  Likelihood fit}
\end{figure}

At the maximum, a Taylor expansion gives $\ln L(a)=\ln L(\hat a)+{1 \over 2} (a-\hat a)^2 {d^2 \ln L \over da^2}$ $\dots$

The maximum likelihood estimator saturates the  MVB, so 
\begin{equation}
V_{\hat a}=-1/\left< d^2 \ln L \over da^2 \right>
\qquad  \sigma_{\hat a}=\sqrt{-{1 \over {d^2 \ln L \over da^2}}}
\quad.
\end{equation}

We approximate the expectation value $\left< d^2 \ln L \over da^2 \right>$ by the actual value in this case
$ \left. d^2 \ln L \over da^2 \right|_{a=\hat a}$ (for a discussion of the introduced inaccuracy, see Ref.~\cite{DeltaML}).

This can be read off the curve, as also shown in Fig.~\ref{fig:ML}. The maximum gives the estimate.
You then draw a line $1\over 2$ below that (of course nowadays this is done within the code, not with pencil and ruler, but the visual image is still valid). This line $\ln L(a)=\ln L(\hat a)-{1 \over 2}$ intersects the likelihood curve at the points
 $a=\hat a \pm \sigma_{\hat a}$.  
 If you are working with $\chi^2$, $L\propto e^{-{1 \over 2}\chi^2}$ so the line is  $\Delta \chi^2=1$.

This gives $\sigma$, or 68\% errors. You can also take $\Delta \ln L=-2$ to get  2 sigma or 95\% errors, or $-4.5$ for 3 sigma errors as desired. For large $N$ these will all be consistent.

\subsection {Combining errors}

Having obtained---by whatever means---errors $\sigma_x, \sigma_y...$ 
how does one combine them to get errors on derived quantities $f(x,y...), g(x,y,...)$?

Suppose $f=Ax+By+C$, with $A,B$ and $C$ constant.
Then it is easy to show that 

\begin{align}
V_f&= \notag
\left< (f - \left< f\right>)^2\right>\\&=  \notag
\left< (Ax+By+C - \left<  Ax+By+C\right>)^2\right> \\
&=  \notag
A^2(\left<x^2\right> - \left<x\right>^2)
+B^2(\left<y^2\right> - \left<y\right>^2)
+2AB(\left<xy\right> - \left<x\right>\left< y \right>)\\
&=A^2 V_x + B^2 V_y + 2AB\, {\rm Cov}_{xy}
\label{eq:COE0}
\quad.
\end{align}

If $f$ is not a simple linear function of $x$ and $y$ then one can use a first order Taylor expansion to
approximate it about a central value $f_0(x_0,y_0)$

\begin{equation}
f(x,y)\approx f_0 
+ \left( {\partial f \over \partial x}\right) (x-x_0)
+ \left( {\partial f \over \partial y}\right) (y-y_0)
\end{equation}

\noindent and application of Eq.~\ref{eq:COE0} gives

\begin{equation}
V_f=
\left( {\partial f \over \partial x}\right)^2 V_x+
\left( {\partial f \over \partial y}\right)^2 V_y+
2 \left( {\partial f \over \partial x}\right)\left( {\partial f \over \partial y}\right) {\rm Cov}_{xy}
\label{eq:COE1}
\end{equation}

\noindent writing the more familiar $\sigma^2$ 
instead of $V$ this is equivalent to

\begin{equation}
\sigma_f^2=
\left( {\partial f \over \partial x}\right)^2 \sigma_x^2+
\left( {\partial f \over \partial y}\right)^2 \sigma_y^2 
+ 2 \rho  \left( {\partial f \over \partial x}\right) \left( {\partial f \over \partial y}\right) \sigma_x \sigma_y
\label{eq:COE2a}
\quad.
\end{equation}

If $x$ and $y$ are independent, which is often but not always the case, this reduces to  what is often known as the
`combination of errors' formula
 
\begin{equation}
\sigma_f^2=
\left( {\partial f \over \partial x}\right)^2 \sigma_x^2+
\left( {\partial f \over \partial y}\right)^2 \sigma_y^2 
\label{eq:COE2}
\quad.
\end{equation}

Extension to more than two variables is trivial: an extra squared term  is added for each and
an extra covariance term for each of the variables (if any) with which it is correlated.

 This can be expressed in language as {\it errors add in quadrature}. This is a friendly fact, as
 the result is smaller than you would get from arithmetic addition. If this puzzles you, it may be helpful to think 
 of this as allowing for the possibility that a positive fluctuation in one variable may be cancelled by a negative fluctuation in
  the other. 
 
There are a couple of special cases we need to consider. 
  If $f$ is a simple product, $f=Axy$, then Eq.~\ref{eq:COE2} gives
  $$\sigma_f^2=(Ay)^2 \sigma_x^2+ (Ax)^2 \sigma_y^2 \ ,$$
  which, dividing by $f^2$, can be written as
  \begin{equation}
  \left({ \sigma_f \over f }\right)^2 =\left( {\sigma_x \over x }\right)^2 +
\left({ \sigma_y \over y }\right)^2.
\label{eq:COE3}
\end{equation}
Furthermore this also applies if $f$ is a simple quotient, 
$f=Ax/y$ or $Ay/x$ or even $A/(xy)$.

This is very elegant, but it should not be overemphasised. Equation~\ref{eq:COE3}
is not fundamental: it only applies in certain cases (products or quotients). Equation~\ref{eq:COE2} is
the fundamental one, and Eq.~\ref{eq:COE3} is just a special case of it.

For example: if you measure the radius of a cylinder as $r=123 \pm 2$ mm and the height as $h=456 \pm 3$ mm
then the volume  $\pi r^2 h$ is $\pi \times 123^2 \times 456 = 21673295 \ {\rm mm}^3$  
with error $\sqrt{(2 \pi r h)^2 \times \sigma_r^2 + ( \pi r^2 )^2 \times \sigma_h^2}=719101$,
so one could write it as $v=(216.73 \pm 0.72) \times 10^5\ {\rm mm}^3$.
The surface area $2 \pi r^2 + 2\pi r h$ is $ 2 \pi \times 123^2 + 2 \pi \times 123 \times 456 = 447470\ {\rm mm}^2$
with error $\sqrt{(4\pi r + 2 \pi h)^2 \sigma_r^2 + (2 \pi r)^2 \sigma_h^2 }= 9121 \ {\rm mm}^2 $---so one could write the result 
as $a=(447.5 \pm 9.1) \times 10^3 \ {\rm mm}^2$.

A full error analysis has to include the treatment of the covariance terms---if only to show that they can be ignored.
Why should the $x$ and $y$ in Eq.~\ref{eq:COE1} be correlated? 
For direct measurements very often (but not always) they will not be.
However the interpretation of results is generally a multistage process. 
From raw numbers of events one computes branching ratios (or cross sections...), from which one computes matrix elements (or particle masses...). Many quantities of interest to theorists are expressed as ratios of experimental numbers. 
And in this interpretation there is plenty of scope for correlations to creep into the analysis.

For example, an experiment might measure a cross section $\sigma(pp \to X) $  from a number of observed events $N$  in the decay channel $X \to \mu^+\mu^-$. One would  
use a formula
$$\sigma={N \over B \eta {\cal L}} \ ,$$
where  $\eta$ is the efficiency for detecting and reconstructing 
an event, $B$ is the branching ratio for $X \to \mu^+\mu^-$, and ${\cal L}$ is the integrated luminosity.
 These will all have errors, and the above prescription can be applied.
 
  However it might also use the $X \to e^+e^-$ channel and then use
 
$$\sigma'={N' \over B' \eta' {\cal L}} \ .$$

Now $\sigma$ and $\sigma'$ are clearly correlated; even though $N$ and $N'$ are 
independent, the same ${\cal L}$ appears in both. If the estimate of ${\cal L}$ is on the high side, that will push both $\sigma$ and $\sigma'$ downwards, and vice versa.  

On the other hand, if a second experiment did the same measurement it would have its own $N$, $\eta$ and ${\cal L}$, but  would be correlated with the first
through using the same branching ratio (taken, presumably, from the Particle Data Group).

To calculate correlations between results we need the equivalent of Eq.~\ref{eq:COE0}

\begin{align}
{\rm Cov}_{fg} &= \notag \left< (f-\langle f \rangle )(g-\langle g \rangle ) \right> \\
&=\left({\partial f \over \partial x} \right) \left( { \partial g \over \partial x} \right) \sigma_x^2
\quad,
\end{align}  

This can all be combined in the general formula which encapsulates all of the ones above

\begin{equation}
{\bf V_f} = {\bf  G V_x \tilde G}
\label{eq:COE4}
\quad,
\end{equation}
where ${\bf V_x}$ is the covariance matrix of the primary quantities (often, as pointed out earlier, this is diagonal),
 ${\bf V_f}$ is the covariance matrix of secondary quantities, and 
 
 \begin{equation}
 G_{ij}={\partial f_i \over \partial x_j}
 \quad.
 \end{equation}
 
 The {\bf G} matrix is rectangular but need not be square. 
 There may be more---or fewer---derived quantities than primary quantities.
 The matrix algebra of ${\bf G}$ and its transpose ${\bf \tilde G}$ 
 ensures that the numbers of rows and columns match for  Eq.~\ref{eq:COE4}.
 
 To show how this works, we go back to our earlier example of a cylinder.
$v$ and $a$ are correlated: if $r$ or $h$ fluctuate upwards (or downwards), that makes both volume and area larger
(or smaller). The matrix ${\bf G}$ is

\begin{equation}
{\bf G}=\left( 
\begin{matrix}
2 \pi r h & \pi r^2 \\
2 \pi (2 r + h) &
2 \pi r
\end{matrix}
\right)
=\left( 
\begin{matrix}
352411 & 47529 \\
4411 & 773
\end{matrix}
\right)
\quad,
\end{equation} 

\noindent the variance matrix $V_x$  is

\begin{equation*}
{\bf V_x}=\left( 
\begin{matrix}
4 & 0\\
0& 9
\end{matrix}
\right)
\end{equation*} 

\noindent and Eq.~\ref{eq:COE4} gives
 
\begin{equation*}
{\bf V_f}=\left( 
\begin{matrix}
517.1 \times 10^9 & 6.548 \times 10^9\\
6.548 \times 10^9 & 83.20 \times 10^6
\end{matrix}
\right)
\end{equation*} 
from which one obtains, as before,
$\sigma_v= 719101, \sigma_a=9121$ but also $\rho=0.9983$.
 
This can be used to provide a useful example of why correlation matters. Suppose
you want to know the volume to surface ratio, $z=v/a$,  of this cylinder. 
Division gives $z=21673295/447470=48.4352$ mm.

If we just use Eq.~\ref{eq:COE2} for the error, this gives $\sigma_z=1.89$ mm. 
Including the correlation term, as in Eq.~\ref{eq:COE2a}, reduces this to
$0.62$ mm---three times smaller. It makes a big difference.

We can also check that this is correct, because the ration ${v \over a}$ can be written as
$\pi r^2 h \over 2 \pi r^2 + 2 \pi r h$, and applying the uncorrelated errors of the original $r$ and $h$ to this also gives
an error of $0.62$ mm.

As a second, hopefully helpful, example we consider a simple straight line fit, $y=mx+c$.
Assuming that all the $N$ $y$ values are measured with the same error $\sigma$,
least squares estimation gives the well known results
\begin{equation}
m={\overline{xy} - \overline x \, \overline y \over \overline{x^2}-{\overline x}^2}
\qquad
c={\overline{y}\, \overline{x^2}  - \overline {xy} \, \overline x \over \overline{x^2}-{\overline x}^2}
\quad.
\end{equation}

For simplicity we write $D=1/(\overline{x^2}-\overline x^2)$.  The differentials are
\begin{equation*}
{\partial m \over \partial y_i}={D \over N} (x_i-\overline x) \qquad
{\partial c \over \partial y_i}={D \over N} (\overline{x^2}- x_i\overline x)
\quad,
\end{equation*}

\noindent from which, remembering that the $y$ values are uncorrelated,

\begin{align*}
V_m=\sigma^2\left({D \over N}\right)^2 \sum (x_i-\overline x)^2=\sigma^2 {D \over N}
 \\
V_c=\sigma^2 \left({D \over N}\right)^2 \sum (\overline{x^2}- x_i\overline x)^2 =\sigma^2 \overline {x^2} {D \over N}\\
{\rm Cov}_{mc}= \sigma^2 \left({D \over N}\right)^2 \sum (x_i-\overline x)(\overline{x^2}-x_i \overline x)=-\sigma^2 \overline x {D \over N}
\end{align*}

\noindent from which the correlation between $m$ and $c$ is just $\rho=- \overline x / \sqrt{\overline{x^2}}$.

This makes sense. Imagine you're fitting a straight line through a set of points with a range of positive $x$ values (so $\overline x$ is positive).   If the rightmost point happened to be a bit higher, that would push the slope $m$ up and the intercept $c$ down. Likewise if the leftmost point happened to be too high that would push the slope down and the intercept up. There is a negative correlation between the two fitted quantities.

 Does it matter?  Sometimes. Not if you're just interested in the slope---or the constant. But suppose you intend to use them to find the expected value of $y$ at some extrapolated $x$. Equation~\ref{eq:COE2a} gives
 \begin{equation*}
 y=m x + c \pm \sqrt {x^2 \sigma_m^2 + \sigma_c^2 + 2 x \rho \sigma_m \sigma_c}
 \end{equation*}
and if, for a typical case where $\overline x$ is positive so $\rho$ is negative, you leave out the correlation term you will overestimate your error.
 
 This is an educational example because this correlation can be avoided.   Shifting to a co-ordinate system in which
  $\overline x$ is zero ensures that the quantities are uncorrelated. This is 
 equivalent to rewriting the well-known $y=mx+c$ formula as $y=m(x-\overline x)+c'$, where
 $m$ is the same as before and $c'=c+m \overline x$. $m$ and $c'$ are now uncorrelated, and 
 error calculations involving them become a lot simpler.

\subsection{Asymmetric errors}

So what happens if you plot the likelihood function and  it is not symmetric like 
Fig.~\ref{fig:ML} but looks more like  Fig.~\ref{fig:asym}?
This  
arises in many cases when numbers are small.  For instance, in a simple Poisson count suppose you observe one event. $P(1;\lambda)=\lambda e^{-\lambda}$ is not symmetric:    $\lambda=1.5$ is more likely to fluctuate down to 1 than $\lambda=0.5$ is to
fluctuate up to 1.

\begin{figure}
\centerline{
\includegraphics[width=6 cm, angle=270, trim={240 0 0 0},clip ] {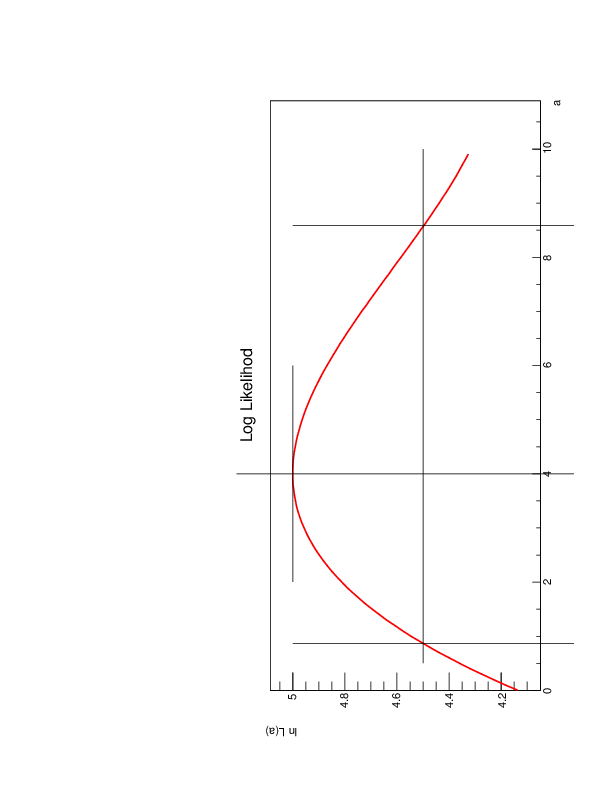}
}
\caption{\label{fig:asym} An asymmetric likelihood curve}
\end{figure}

You can read off $\sigma_+$ and $\sigma_-$ from the two $\Delta \ln L=-{1 \over 2}$ crossings, but they are different.
The result can then be given as $a^{+\sigma_+}_{-\sigma_-}$. What happens after that?

The first advice is to avoid this if possible.  
If you get $\hat a=4.56$ with $\sigma_+=1.61, \sigma_-=1.59$ 
 then quote this as $4.6 \pm 1.6$ rather than $4.56^{+1.61}_{-1.59}$.
 Those extra significant digits have no real meaning. If you can convince yourself that the difference between
 $\sigma_+$ and $\sigma_-$ is small enough to be ignored then you should do so, as the alternative brings in a whole 
 lot of trouble and it's not worth it.
 
 But there will be some cases where the difference is too great to be swept away, so
 let's consider that case.
 There are two problems that arise: combination of measurements and combination of errors.
 
 \subsubsection{Combination of measurements with asymmetric errors}
 
Suppose you have two measurements of the same parameter $a$: $\hat {a_1}^{+\sigma^+_1}_{-\sigma^-_1}$
and $\hat {a_2}^{+\sigma^+_2}_{-\sigma^-_2}$ and you want to combine them to give the best estimate and, of course, its error. For symmetric errors the answer is well established to be
$\hat a = {\hat a_1/\sigma_1^2 + \hat a_2/\sigma_2^2 \over 1/\sigma_1^2 + 1/\sigma_2^2}$.

If you know the likelihood functions, you can do it. The joint likelihood is just the sum.
This is shown in Fig.~\ref{fig:combine1} where 
the
 red  and green curves  are measurements of $a$. 
The log likelihood functions just add (blue), from which the peak is found and the $\Delta \ln L=-\half$ errors read off.

\begin{figure}
[h]
\centerline{
\includegraphics[width=6 cm, angle=270, trim={240 0 0 0},clip ] {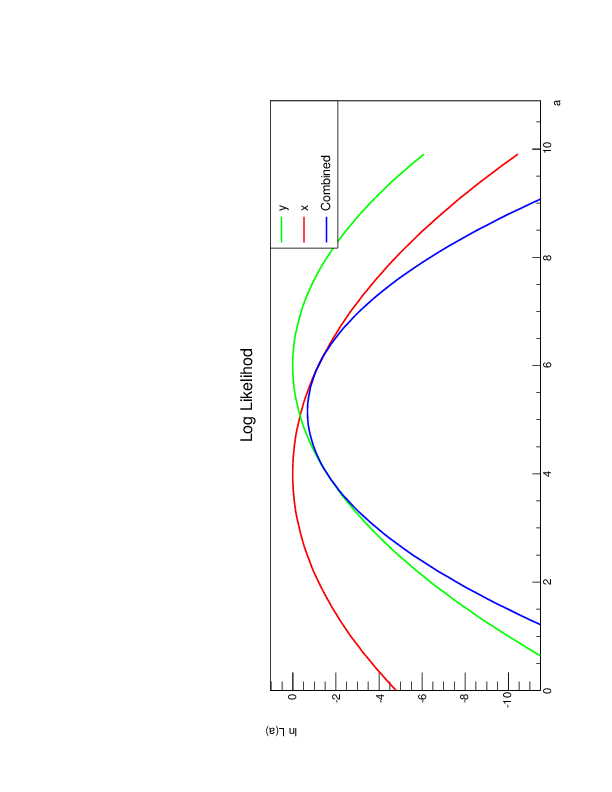}}
\caption{\label{fig:combine1} Combination of two likelihood functions (red and green) to give the total (blue)}
\end{figure}

 But you don't know the full likelihood function: just 3 points (and that it had a maximum at the second).
 There are, of course, an infinite number of curves that could be drawn, and several models have been tried (cubics, constrained quartic...) on likely instances---see Ref.~\cite{asym} for details. Some do better than others.
The two most plausible  are

\begin{equation}
\ln L = -{1 \over 2}\left( {a-\hat a \over \sigma+\sigma'(a-\hat a)}\right)^2
 \quad \text{and}
\end{equation}
\begin{equation}
 \ln L = -{1 \over 2} {\left( a-\hat a \right)^2 \over V+V'(a-\hat a)}
 \quad.
 \end{equation}

These are similar to the Gaussian parabola, but the denominator is not constant. It varies with the value of $a$, being linear either in the standard deviation or in the variance.
Both are pretty good.  The first does better with errors on $\log a$ (which are asymmetric if $a$ is symmetric: such asymmetric error bars are often seen on plots where the $y$ axis is logarithmic), the
second does better with Poisson measurements.

 From the 3 numbers given one readily obtains
 \begin{equation}
 \sigma={2 \sigma^+\sigma^- \over \sigma^+ + \sigma^-} \qquad 
 \sigma'={\sigma^+-\sigma^- \over \sigma^+ + \sigma^-}
 \label{eq:asyms}
 \end{equation}
 or, if preferred
 \begin{equation}
 V= \sigma^+\sigma^- \qquad
 V'=\sigma^+-\sigma^- 
 \label{eq:asymV}
 \quad.
 \end{equation}
 
From the total likelihood you then find the maximum of sum, numerically, and the $\Delta \ln L=-{1\over 2}$ points.

Code for doing this is available on GitHub\footnote{\url{ https://github.com/RogerJBarlow/Asymmetric-Errors}} in
both R and Root.

\begin{figure}[h]
\begin{center}
\includegraphics[width=6 cm, angle=270,trim={260 0 0 100},clip] {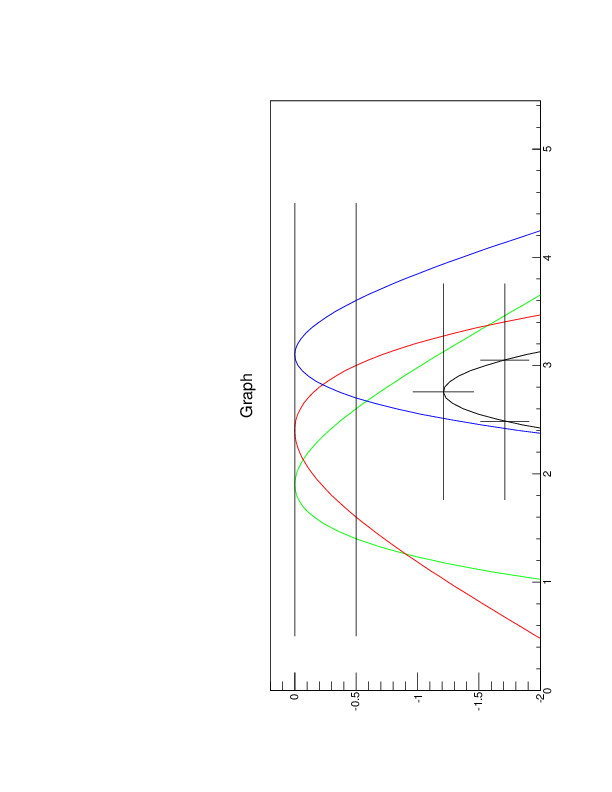}
\caption{\label{fig:asymex} Combining three asymmetric measurements}
\end{center}
\end{figure}

An example is shown in Fig.~\ref{fig:asymex}. Combining $1.9^{+0.7}_{-0.5}$, $2.4^{+0.6}_{-0.8}$ and $3.1^{+0.5}_{-0.4}$ gives $2.76 ^{+0.29}_{-0.27} \ .$

 \subsubsection{Combination of errors for asymmetric errors}

For symmetric errors, given $x  \pm \sigma_x, y \pm \sigma_y$, (and $\rho_{xy}=0$)  the error on 
$f(x,y)$
is  the sum in quadrature: $\sigma_f^2 =\left( {\partial  f \over \partial x}\right)^2  \sigma_x^2 + \left( {\partial  f \over \partial y}\right)^2
\sigma_y^2$.
 What is  the equivalent for the error on $f(x,y)$ when the errors are asymmetric,
 $x ^{+\sigma^+_x}_{-\sigma^-_x}, y^{+\sigma^+_y}_{-\sigma^-_y}$? Such a problem arises frequently at the end of an analysis when the systematic errors from various sources are all combined.
 
 The standard procedure---which you will see done, though it has not, to my knowledge, been written down anywhere---is to add the positive and negative errors
 in quadrature separately:  ${\sigma^+_f}^2={\sigma^+_x}^2+{\sigma^+_y}^2$,\quad 
${\sigma^-_f}^2={\sigma^-_x}^2+{\sigma^-_y}^2$.
This looks plausible, but it is 
 {\em manifestly wrong} as it breaks the central limit theorem.

To see this, suppose you have to average 
 $N$ i.i.d. variables each with the same errors which are asymmetric:  $\sigma^+ = 2 \sigma^-$ . 
The 
standard procedure reduces both $\sigma^+$ and $\sigma^-$ by a factor $1/\sqrt N$,  but the skewness remains.
The positive error is twice the negative error.  This is therefore not Gaussian, and never will be, even as $N \to \infty$.
 
 You can see what's happening by considering the combination of two of these measurements.  They both may fluctuate upwards, or they both may fluctuate downwards, and yes, the upward fluctuation will be, on average, twice as big.  But there is a 50\%
 chance of one upward and one downward fluctuation, which is not considered in the standard procedure.
 
 For simplicity we write $z_i={\partial f \over \partial x_i} (x_i-x^0_i)$, the deviation of the parameter 
 from its nominal value, scaled by the differential. The individual likelihoods are  again parametrized
 as Gaussian with a linear dependence of the standard deviation or of the variance, giving
 
 \begin{equation}
 \ln L(\vec z)=- \half \sum_i \left( { z_i \over \sigma_i + \sigma'_i z_i}\right)^2 \qquad {\rm or}\qquad
 - \half \sum_i  { z_i ^2\over V_i + V'_i z_i}
 \quad,
 \end{equation}
 where $\sigma, \sigma', V,V'$ are obtained from Eqs.~\ref{eq:asyms} or \ref{eq:asymV}.
 
 The $z_i$ are nuisance parameters (as described later) and can be removed by profiling. 
 Let $u=\sum z_i$ be the total deviation in the quoted $f$ arising from the individual deviations.
 We form $\hat L(u)$ as the maximum of $L(\vec z)$ subject to the constraint $\sum_i z_i=u$.
 The method of undetermined multipliers readily gives the solution
 \begin{equation}
 \label{eq:comsol1}
 z_i=u {w_i \over \sum_j w_j}
 \quad,
 \end{equation}
 where
  \begin{equation}
 \label{eq:comsol2}
 w_i = {(\sigma_i  +  \sigma'_i z_i)^3 \over 2 \sigma_i}
 \qquad {\rm or } \qquad
 {(V_i+V'_i z_i)^2 \over 2V_i + V'_i z_i}
 \quad.
 \end{equation}
 The equations are nonlinear, but can be solved iteratively. At $u=0$ all the $z_i$ are zero. Increasing (or decreasing) $u$ in small steps, Eqs.~\ref{eq:comsol1} and \ref{eq:comsol2} are applied successively to give the  $z_i$ and the $w_i$: convergence is rapid.   The value of $u$ which maximises the likelihood should in principle be applied as a correction to the quoted result.
 
 Programs to do this are also available on the GitHub site.
 
 As an example, consider a counting experiment with a number of backgrounds, each determined by an ancillary Poisson experiment, and that for simplicity each background was determined  by running the apparatus for the same time as the actual experiment. (In practice this is unlikely, but scale factors
 can easily be added.)
 
 Suppose two backgrounds are measured, one giving four events and the other five. These would be reported, using $\Delta ln L=-\half$ errors, as $4^{+2.346}_{-1.682}$ and $5^{+2.581}_{-1.916}$.
 The method, using linear $V$, gives the combined error on the background count as ${}^{+3.333}_{-2.668}$.
 
 In this simple case we can check the result against the total background count of nine events, which has errors ${}^{+3.342}_{-2.676}$. The agreement is impressive.  Further examples of the same total, partitioned differently, are shown in table~\ref{tab:asymcom}.
 
 \begin{table}[h]
 \begin{centering}
 \begin{tabular}{ l c c c c }
 Inputs & \multicolumn {2} {c}  {Linear $\sigma$}& \multicolumn {2} {c}  {Linear $V$}\\
 & $\sigma^-$ & $\sigma^+$ & $\sigma^-$ & $\sigma^+$ \\
 \hline
 4+5 & 2.653 & 3.310 & 2.668 & 3.333 \\
  3+6 & 2.653 & 3.310 & 2.668 & 3.333 \\
   2+7 & 2.653 & 3.310 & 2.668 & 3.333 \\
   2+7 & 2.653 & 3.310 & 2.668 & 3.333 \\
    3+3+3 & 2.630 & 3.278 & 2.659 & 3.323 \\
     1+1+1+1+1+1+1+1+1 & 2.500 & 3.098 & 2.610 & 3.270 \\
 \end{tabular}
 \caption{\label{tab:asymcom} Various combinations of Poisson errors. The target value is $\sigma^-=2.676$, $\sigma^+=3.342$}
 \end{centering}
 \end{table}
 
  \subsection{Errors in 2 or more dimensions}

For 2 (or more) dimensions, one plots the log likelihood and defines regions using contours in $\Delta \ln L$ (or $\Delta \chi^2\equiv - 2 \Delta \ln L$). An example is given in Fig.~\ref{fig:CMS}.

\begin{figure}[h]
\begin{centering}
\includegraphics[width=10 cm, angle=0, trim={0 0 0 0},clip ] {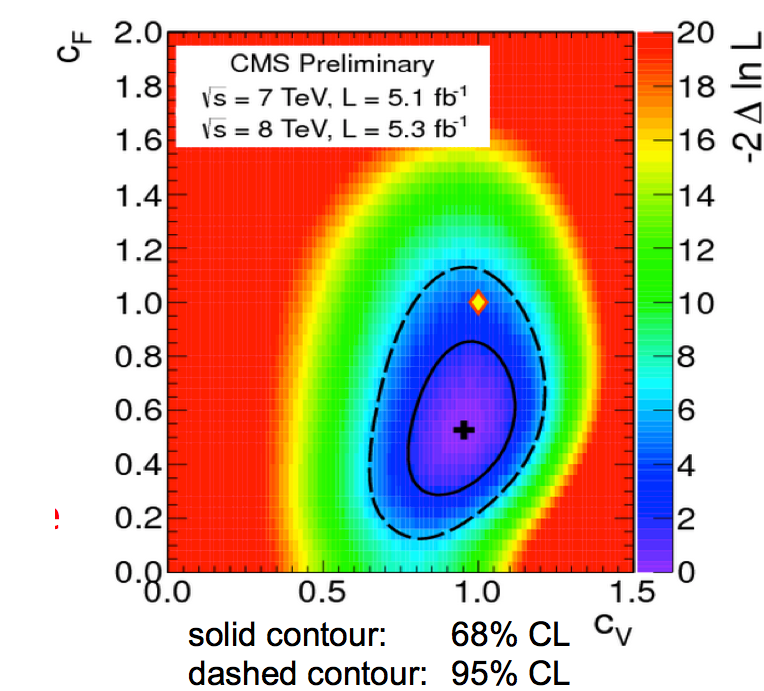}

\caption{\label{fig:CMS} CMS results on $C_V$ and $C_F$, taken from~Ref.~\cite{CMS}}
\end{centering}
\end{figure}

The link between the $\Delta \ln L$ values and the significance changes. 
In 1D, there is a 68\% probability of a measurement falling within 1 $\sigma$.
In 2D, a  1$\sigma$ square would give a probability $0.68^2 = 47\%$. 
If one rounds off the corners and draws a 1$\sigma$ contour 
at $\Delta \ln L=-\half$ this falls to 39\%. 
To retrieve the full 68\% one has to draw a contour at $\Delta \ln L=-1.14$, or equivalently 
  $\Delta \chi^2=2.27$.
For 95\%  use  $\Delta \chi^2=5.99$ or $\Delta \ln L=-3.00$.

The necessary value is obtained from the $\chi^2$ distribution---described later. It can be found by the R function {\tt qchisq(p,n)} or the Root function {\tt TMath::ChiSquareQuantile(p,n)}, where the desired 
probability {\tt p}  and number of degrees of freedom {\tt n} are the arguments given.

\subsubsection{Nuisance parameters}

In the example of Fig.~\ref{fig:CMS},  both $C_V$ and $C_F$ are interesting.
But in many cases one is interested only in one (or some) of the quantities and the others
are `nuisance parameters' that one would like to remove, reducing the dimensionality of the 
quoted result.   There are two methods of doing this, one (basically) frequentist and one Bayesian.

The frequentist uses the {\it profile likelihood} technique. Suppose that there are two parameters,
$a_1$ and $a_2$, where $a_2$ is a nuisance parameter, and so one wants to reduce the 
joint likelihood function $L(x;a_1,a_2)$ to some function $\hat L(a_1)$.
To do this one scans across the values of $a_1$ and inserts $\hat{\hat a}_2(a_1)$, the value of $a_2$   which maximises the likelihood for that particular $a_1$

\begin{equation}
\hat L (x,a_1)=L(a_1,\hat{\hat a}_2(a_1))
\end{equation}

\noindent and the location of the maximum and the $\Delta \ln L=\-\half$ errors are
read off as usual. 

To see why this works---though this is not a very rigorous motivation---suppose one had a likelihood function
as shown in Fig.~\ref{fig:profile}.

\begin{figure}[h]
\begin{centering}
\includegraphics[width=7 cm, angle=270, trim={0 0 0 0},clip ] {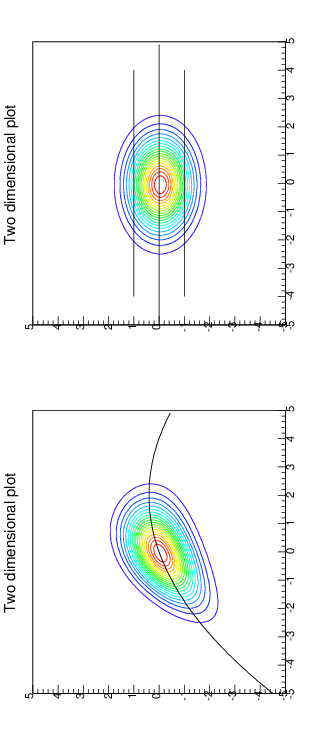}
\caption{\label{fig:profile} Justification of the likelihood profile method}
\end{centering}
\end{figure}

 The horizontal axis is for the parameter of interest, $a_1$, and the vertical for the nuisance parameter $a_2$.
  
Different values of $a_2$ give different results (central and errors) for $a_1$.

If it is possible to transform to $a_2'(a_1,a_2)$ so that $L$ factorises, then we can write
$L(a_1,a_2')=L_1(a_1)L_2(a_2')$: this is shown in the plot on the right.
We suppose that this is indeed possible. In the case here, and other not-too-complicated cases, it clearly is, although
it will not be so in more complicated topologies with multiple peaks. 

Then using the transformed graph, 
whatever the value of $a_2'$, one would get the same result for $a_1$.
Then one can present this result for $a_1$, independent of anything about $a_2'$.

There is no need to factorise explicitly: the
path of central $a_2'$ value as a function of $a_1$ (the central of the 3 lines on the right hand plot)
is the path of the peak, and that path can be located in the first plot (the transformation only stretches the $a_2$ axis, it does not change the heights).
 
 The Bayesian method uses the technique called {\it marginalisation}, which 
 just integrates over $a_2$.
Frequentists can not do this as they are not allowed to integrate likelihoods over the parameter:
$\int P(x;a)\, dx$ is fine, but $\int P(x;a)\, da$ is off limits.
 Nevertheless this can be a very helpful alternative to profiling, specially for many nuisance parameters.
But if you use it you must be aware that this is strictly Bayesian.
Reparametrizing $a_2$ (or choosing a different prior) will give different results for $a_1$.
In many cases, where the effect of the nuisance parameter is small, this does not
have a big effect on the result.
 
 \subsection{Systematic errors}
  
This can be a touchy subject.  There is a lot of bad practice out there. Muddled thinking and following traditional
 procedures without understanding. 
  When statistical errors dominated, this didn't matter much.  In the days of particle factories and big data samples, it does. 
 
 \subsubsection{What is a systematic error?}
 
 Consider these two quotations, from eminent and widely-read authorities.
 
 R. Bevington defines

\begin{quotation}
`Systematic error: 
reproducible inaccuracy introduced by faulty equipment, calibration, or technique.'~\cite{Bevington},
\end{quotation}

\noindent whereas J. Orear writes

 \begin{quotation}
 `Systematic effects is a general category which includes effects such as background, scanning efficiency, energy resolution, variation of counter efficiency with beam position, and energy, dead time, etc. The uncertainty in the estimation of such a systematic effect is called a systematic error.'~\cite{Orear}.
 
\end{quotation}
 
Read these carefully and you will see that they are contradictory. They are not talking about the same thing.  Furthermore, Orear is {RIGHT} and
Bevington is {WRONG}---as are a lot of other books and websites.

 We teach undergraduates the difference between measurement {\it errors}, which are
 part of doing science, and {\em mistakes}. They are not the same. 
 If you measure a potential of 12.3 V as 12.4 V, with a voltmeter 
 accurate to 0.1V, that is fine.  Even if you measure 12.5 V.
 If you measure it as 124 V, that is a mistake.
 
 In the quotes above, Bevington is describing {\it systematic mistakes} (the word `faulty' is the key)
 whereas 
 Orear is describing {\it systematic uncertainties}---which are `errors' in the 
 way we use the term.

 There is a case for saying one should avoid the term  `systematic error' and always use `uncertainty' or 'mistake'. This is 
 probably impossible. But you should always know which you mean.
  
  Restricting ourselves to uncertainties (we will come back to mistakes later) here are some typical examples:
  
  \begin{itemize}
 
 \item
 Track momenta from $p_i=0.3 B \rho_i$  have statistical errors from $\rho$ and systematic errors  from $B$,
 
 \item
 Calorimeter energies from $E_i=\alpha D_i + \beta$ have statistical errors from the digitised
 light signal $D_i$ and systematic errors from the calibration $\alpha,\beta$, and
 
 \item Branching ratios from $Br={N_D - B \over \eta N_T}$ have statistical errors
 from $N_D$ and systematic errors from efficiency $\eta$, background $B$, total $N_T$\ .
 
 \end{itemize}
 
 Systematic uncertainties can be either Bayesian or Frequentist.
There are clearly frequentist cases where errors have been  determined by an {\em ancillary experiment}
(real or simulated), such as 
 magnetic field measurements, calorimeter calibration in a testbeam,  and efficiencies from Monte Carlo simulations.
 (Sometimes the ancillary experiment is also the main experiment---e.g. in estimating background from sidebands.)
 There are also uncertainties that can only be Bayesian, e.g.
 when a  theorist tells you that their calculation is good to 5\% (or whatever) or an
 experimentalist affirms that the calibration will not have shifted during the run by more than
 2\% (or whatever).
  
 \subsubsection{How to handle them: correlations}
 
 Working with systematic errors is actually quite straightforward. They obey the same rules as statistical uncertainties.
  
   We write  $x=12.2 \pm 0.3 \pm 0.4$ `where the first error is statistical and the second is systematic', but it would be valid to write $x=12.2 \pm 0.5$.
  For single measurement the extra information given by the two separate numbers is small.  (In this case it just tells you that there is little to be gained by 
  increasing the size of the data sample).
 For multiple measurements e.g. $x_a=12.2 \pm 0.3, x_b=17.1 \pm 0.4,  all  \pm 0.5$
 the
  extra information is important, as results are correlated.
 Such cases arise, for example, in cross section measurements with  a common luminosity error, or branching ratios with common efficiency.
 
 Such a correlation means that  taking more measurements and averaging does not reduce the error.
 Also there is no way to estimate $\sigma_{sys}$ from the data---hence no check on the goodness of fit  from a  $\chi^2$ test.
 
 \subsubsection{Handling systematic errors in your analysis}
 
It is useful to consider systematic errors as having three types:

\begin{enumerate}

\item Uncertainty in an explicit continuous parameter.
For example an
  uncertainty in efficiency, background and luminosity in determining a  branching ratio or cross section.
 For these the 
 standard combination of errors formula and algebra are usable, just like undergraduate labs.
  
\item
Uncertainty in an implicit continuous parameter.
For example: MC tuning parameters ($\sigma_{p_T}$, polarisation $\dots$).
These are not amenable to algebra. Instead one calculates the result for different parameter values, typically at $\pm \sigma$, and observes the variation in the result, as illustrated in Fig.~\ref{fig:MCsys}.

\begin{figure}[h]
\begin{centering}
 \includegraphics[width=8 cm]{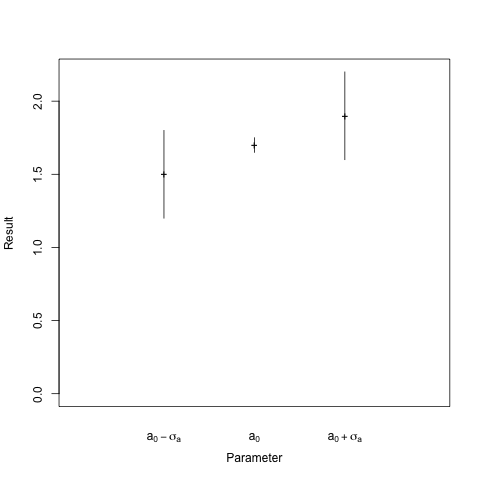}
 \caption{\label{fig:MCsys} Evaluating the effect of an implicit systematic uncertainty}
 \end{centering}
\end{figure}
 
 Hopefully the effect is equal but opposite---if not then one can reluctantly quote an asymmetric error. 
Also your analysis results will have errors due to finite MC statistics.
Some people add these in quadrature. This is wrong.
The technically correct thing to do is to subtract them in quadrature, but this is not advised.
 
 \item Discrete uncertainties:
 
 These typically occur  in model choices. Using a different Monte Carlo for background---or signal---gives you a (slightly) different result. How do you include this uncertainty?
 
 The situation depends on the status of the models.  Sometimes one is preferred, sometimes they are 
 all equal (more or less).
 
 With 1 preferred model and one other, quote $R_1 \pm |R_1-R_2|$\ .
 
 With 2 models of equal status, quote ${R_1+R_2 \over 2} \pm |{R_1-R_2 \over \sqrt 2}|$\ .
 
 With N models: take $\overline R  \pm \sqrt{{{N \over N-1}(\overline R^2}-{\overline R}^2)}$
 or similar mean value.
 
 2 extreme models: take  ${R_1+R_2 \over 2} \pm{|R_1-R_2| \over \sqrt{12}}$\ .
 
 These are just ballpark estimates. Do not push them too hard. 
 If the difference is not small, you have a problem---which can be an opportunity
 to study model differences. 

\end{enumerate}

\subsubsection{Checking the analysis}
   
{\em ``As we know, there are known knowns. There are things we know that we know. There are known unknowns. That is to say, there are things that we know we don't know. But there are also unknown unknowns. There are things we don't know we don't know."}

\rightline{Donald H. Rumsfeld}
 
 Errors are not mistakes---but mistakes still happen.
Statistical tools can help find them.
 Check your result by repeating the analysis with changes which {\it should} make no difference:
 
 \begin{itemize}
 \item Data subsets,
 \item Magnet up/down,
 \item Different selection cuts,
 \item Changing histogram bin size and fit ranges,
 \item Changing parametrization (including order of polynomial),
 \item Changing fit technique,
 \item Looking for impossibilities,
 \item $\dots$
 \end{itemize}
 
 The more tests the better. You cannot prove the analysis is correct. But the more tests it survives the
 more likely your colleagues\footnote{and eventually even you}  will be to believe the result.
 
 For example: in the paper reporting the first measurement of CP violation in $B$ mesons the BaBar Collaboration~\cite{BaBarCP} reported
 
 \begin{quotation}
 `$\dots$ consistency checks, including separation of the decay by decay mode,
 tagging category and $B_{tag}$ flavour $\dots$  We also fit the samples of non-CP decay modes for $\sin 2\beta$ with no statistically significant difference found.'
 \end{quotation}
 
 If your analysis passes a test  then  {\it tick the box and move on}.
 Do not add the discrepancy to the systematic error. Many people do---and your supervisor and your review committee may want you to do so. Do not give in.
 
 \begin{itemize}
 
 \item It's illogical,
 
 \item It penalises diligence, and
 
 \item Errors get inflated.
 
 \end{itemize}

 If your analysis fails a test then 
worry! 

 \begin{itemize}
 
 \item Check the test. Very often this turns out to be faulty.
 
 \item Check the analysis. Find mistake, enjoy improvement.
 
 \item Worry. Consider whether the effect might be real. (E.g. June's results are different
 from July's.  Temperature effect? If so can (i) compensate and (ii) introduce
 implicit systematic uncertainty).  

 \item Worry harder. Ask colleagues, look at other experiments.
  
 \end{itemize}

 Only as a last resort, add the term to the systematic error. Remember that this
 could be a hint of something much bigger and nastier.
\newpage
    \subsubsection{Clearing up a possible confusion}
 
 What's the difference between?
 
Evaluating implicit systematic errors: vary lots of parameters, see what happens to the result, and include in systematic error.
 
Checks:  vary lots of parameters, see what happens to the result, and don't include in systematic error.
 
 If you find yourself in such a situation there are actually two  ways to tell the difference.
 
 (1) Are you expecting to see an effect?  If so, it's an evaluation, if not, it's a check.
 
 (2) Do you clearly know how much to vary them by? If so, it's an evaluation. If not,
 it's a check.
 
 These cover even complicated cases such as a trigger energy cut where the energy calibration is uncertain---and it 
 may be simpler to simulate the effect by varying the cut rather than the calibration.
 
  \subsubsection{So finally:}
 
 \begin{enumerate}
 
 \item Thou shalt never say `systematic error' when thou meanest
`systematic effect' or `systematic mistake'.

\item Thou shalt know at all times 
whether what thou performest is a check for a mistake
or an evaluation  of an uncertainty.

\item Thou shalt not incorporate successful check results
into thy total systematic error and make thereby a shield
to hide thy dodgy result.

\item Thou shalt not incorporate failed check results unless thou art
truly at thy wits' end.

\item Thou shalt not add uncertainties on uncertainties in quadrature.
If they are larger than chickenfeed thou shalt generate more
Monte Carlo until they shrink.

\item Thou shalt say what thou doest, and thou shalt be able
to justify it out of thine own mouth; not the mouth of
thy supervisor, nor thy colleague who did the analysis last time,
nor thy local statistics guru, nor thy mate down the pub.

\end{enumerate}

\rightline{Do these, and thou shalt flourish, and thine analysis likewise.}
  
\section{Goodness of fit}

You have the best fit model to your data---but is it good enough? The upper plot in Fig.~\ref{fig:badfit}
shows the best straight line through a set of points which are clearly not well described by a straight line. How can one quantify this?

You construct some measure of agreement---call  it  $t$---between the model and the data.
Convention: $t\geq 0$, $t=0$ is perfect agreement. Worse agreement implies larger $t$.
The null hypothesis $H_0$ is that the model did indeed produce this data.
You calculate the
$p-$value: the probability under $H_0$ of getting a $t$ this bad, or worse. This is shown schematically in the lower plot.
Usually this can be done using known algebra---if not one can use simulation (a so-called `Toy Monte Carlo').
 
 \begin{figure}[h]
 \begin{centering}
\includegraphics[width=8 cm, angle=0, trim={0 0 0 0},clip ] {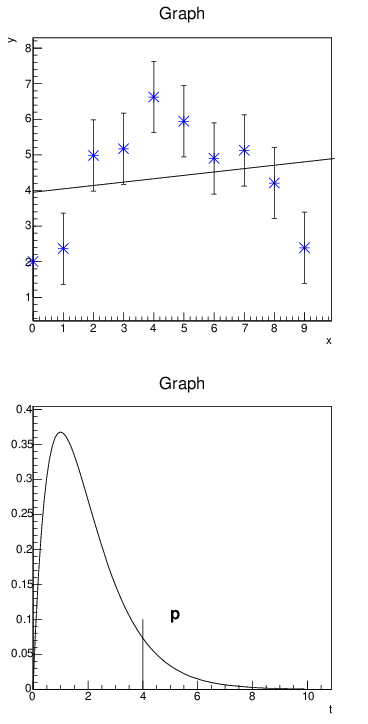}
\caption{\label{fig:badfit} The best fit to the data may not be good enough}
\end{centering}
\end{figure}
 
\subsection{\texorpdfstring
{The $\chi^2$ distribution}
{The chi distribution}}
The overwhelmingly most used such measure of agreement is the quantity $\chi^2$
\begin{equation}
\chi^2 = \sum_1^N \left({y_i-f(x_i) \over \sigma_i}\right)^2
\label{eq:chisq}
\quad.
\end{equation}
 In words: the total of the squared differences between prediction and data, scaled by the expected error.  
 Obviously each term will be about 1, so $\left<\chi^2\right> \approx N$,
and this turns out to be exact.

The distribution for $\chi^2$ is given by
 \begin{equation}
P(\chi^2;N)={1 \over 2^{N/2} \Gamma(N/2)} \chi^{N-2} e^{-\chi^2/2} 
\end{equation}
 shown in Fig.~\ref{fig:chisq1}, 
though this is in fact not much used: one is usually interested in the $p-$value,
the probability (under the null hypothesis) of getting a value of $\chi^2$ as large as, or larger than,  the one  observed. This can be found  in ROOT with {\tt TMath::Prob(chisquared,ndf)},
and
 in R from  {\tt 1-pchisq(chisquared,ndf)}.
\begin{figure}[h]
\begin{centering}
 \includegraphics[width=8 cm, angle=270, trim={10 0 0 0},clip ] {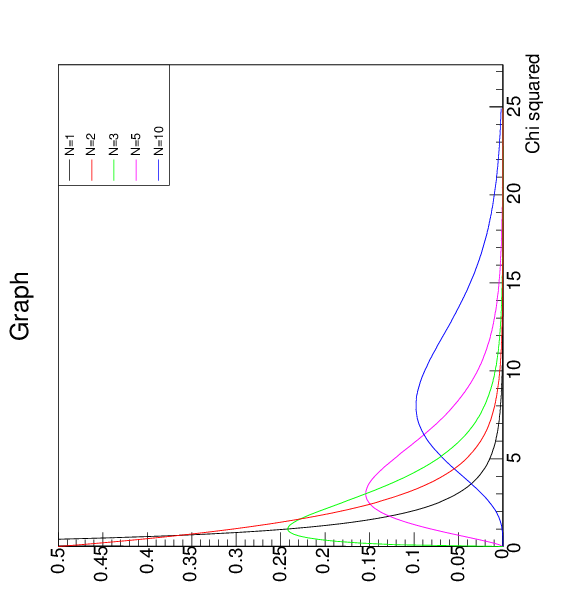}
 \caption{\label{fig:chisq1} The $\chi^2$ distribution for various $N$}
\end{centering}
 \end{figure}
 
Thus for example with 
  $N=10,\chi^2=15$ then $p=0.13$. This is probably OK. 
 But for
 $N=10,\chi^2=20$ then $p=0.03$, which is probably not OK.
 
If the model has parameters which have been adjusted to fit the data, this 
 clearly reduces $\chi^2$. It is a very useful fact that 
the result also follows a $\chi^2$ distribution for $NDF=N_{data}-N_{parameters}$
where $NDF$ is called the `number of   degrees of freedom'.
 
If your $\chi^2$ is suspiciously big, there are 4 possible reasons:
\begin{enumerate}
\item Your model is wrong,
\item Your data are wrong,
\item Your errors are too small, or
\item You are unlucky.
\end{enumerate}

If your $\chi^2$ is suspiciously small there are 2 possible reasons:
\begin{enumerate}
\item Your errors are too big, or
\item You are lucky.
\end{enumerate}

\subsection{Wilks' theorem}
 
The Likelihood on its own tells you {\it nothing}.
Even if you include all the constant factors normally omitted in maximisation.
This may seem counter-intuitive, but it is inescapably true.

There is a theorem due to 
Wilks which is frequently invoked and appears to  link   likelihood and $\chi^2$,
but it does so only in very specific circumstances. 
 Given two nested models, for large $N$
the improvement in $ \ln L$ is distributed like $\chi^2$ in $- 2\Delta  \ln L$, with $NDF$ the number of extra parameters.

So suppose you have some data with many $(x,y)$ values and two models,  Model 1 being linear and  Model 2 quadratic.
You maximise the likelihood using  Model 1 and then using  Model 2: the Likelihood increases as more parameters are available ($NDF=1$). If this increase is significantly 
more than $N$ that justifies using Model 2 rather than Model 1. 
So it may tell you whether or not the extra term in a quadratic gives a meaningful improvement, but not 
whether the final quadratic (or linear) model is a good one.

Even this has an important exception. it does 
 not apply if Model 2 contains a parameter which is meaningless under Model 1. 
This is a surprisingly  common occurrence. Model 1 may be background, Model 2  background  plus a Breit-Wigner  with adjustable mass, width and normalization ($NDF=3$).
The mass and the width are meaningless under Model 1 so Wilks' theorem does not apply and the improvement in likelihood cannot be translated into a $\chi^2$ for testing.
 
 \subsection{Toy Monte Carlos and likelihood for goodness of fit}
 
Although the likelihood contains no information about the goodness of fit of the model,
an obvious way to get such information is to  run many simulations of the model, plot the spread of fitted likelihoods and use it to get the $p-$value.

This may be obvious, but it is wrong~\cite{Heinrich}.
Consider a test case observing decay times where the model is 
a simple exponential $P(t)={ 1 \over \tau}e^{-t/\tau}$, with $\tau$ an adjustable parameter.
Then
 you get the 
Log Likelihood $\sum (-t_i/\tau - \ln \tau)=-N(\overline t /\tau + \ln \tau)$
and maximum likelihood  gives $\hat t = \overline t = {1 \over N} \sum_i t_i$,
so
$\ln L(\hat t;x)= - N(1 + \ln \overline t)$ . This holds
whatever the original sample $\{t_i\}$  looks like:
 any distribution with the same $\overline t$ has the same likelihood, after fitting.

\section{Upper limits}

Many analyses are `searches for...' and 
most of these are unsuccessful.
But you have to say something!  Not just `We looked, but we didn't see anything'.
This is done using the construction of frequentist confidence intervals and/or Bayesian credible intervals.

\subsection{Frequentist confidence}

\label{sec:confidence}
 
Going back to the discussion of the basics, for frequentists  the probability that it will rain tomorrow is meaningless:
there is only one tomorrow, it will either rain or it will not, there is no
ensemble.
The probability $N_\mathrm{rain}/N_\mathrm{tomorrows}$ is either 0 or 1.
To talk about 
$P_\mathrm{rain}$ is "unscientific"~\cite{vonMises}.

This is unhelpful. But there is a workaround.
 
Suppose some forecast says it will rain and
studies show this forecast is correct 90\% of the time.
We now have an ensemble of statements, and can say:
`The statement `It will rain tomorrow' has a 90\% probability of being true'.
We shorten this to  `It will rain tomorrow, with 90\% confidence'.
We state X with confidence $P$ if X is a member of an ensemble of statements of which at least $P$ are true.

Note the `at least' which has crept into the definition. There are two reasons for it:
\begin{enumerate}
\item Higher confidences embrace lower ones. If X at 95\% then X at 90\%, and
\item We can cater for composite hypotheses which are not completely defined.
\end{enumerate}

 The familiar quoted error is in fact a confidence statement. Consider as an illustration
 the Higgs mass measurement (current at the time of writing)
 $M_H=125.09 \pm 0.24$~GeV.
 This does not mean that the probability of the Higgs mass being in the range 
 $124.85 < M_H < 125.33$~GeV  is 68\%: 
 the Higgs mass is a single, unique, number which either lies in this interval or it does not. 
 What we are saying is that 
$M_H$ has been measured to be $125.09$~GeV with a technique that will give a value within 0.24~GeV of the true value 68\% of the time.
We say:  $124.85 < M_H < 125.33~GeV$ with 68\% confidence.
The statement is either true or false (time will tell), but it belongs to a collection of statements of which (at least) 68\% are true.

So we construct 
  {\it confidence regions} {also known as confidence intervals}
 $[x_-,x_+]$ such that $\int_{x_-}^{x_+} P(x) \, dx=CL$.
We have not only a
choice of the probability content (68\%, 90\%, 95\%, 99\%...) to work with but also of strategy. Common options are:

\begin{enumerate}
\item Symmetric: $\hat x-x_- = x_+ - \hat x$\ ,
\item Shortest: Interval that minimises $x_+-x_-$\ ,
\item Central: $\int_{-\infty}^{x_-}P(x)\, dx=\int_{x_+}^\infty P(x)\, dx = {1 \over 2} (1-CL)$\ ,
\item Upper Limit: $x_-=-\infty$, $\int_{x_+}^\infty P(x)\ , dx=1-CL$\ , and
\item Lower Limit:  $x_+=\infty$, $\int_{-\infty}^{x_-} P(x)\ , dx=1-CL$\ .
\end{enumerate}

For the Gaussian (or any symmetric PDF) 1-3 are the same. 

We are particularly concerned with the upper limit: 
 we observe some small value $x$. We find a value $x_+$ such that for values of $x_+$ or more the probability of getting a result as small as $x$, or even less, is $1-CL$, or even less.

\subsection{Confidence belts}
 
We have shown that a simple Gaussian measurement is basically a statement about 
confidence regions. 
 $x=100\pm 10 $ implies that  [90,110] is the 68\% central confidence region.
 
 We want to extend this to less simple scenarios. As a first step, we consider a proportional Gaussian.
 Suppose we measure $x=100$ from Gaussian measurement  with $\sigma= 0.1 x$  (a 10\% measurement---which is realistic).
If the true value is 90 the error is $\sigma=9$ so $x=100$ is more than one standard deviation, whereas if the true value is 110 then $\sigma=11$ and it is less than one standard deviation. 90 and 110 are not equidistant from 100.

This is done with a technique called a confidence belt. The key point is that they are
 are constructed horizontally and read vertically, using the following procedure (as shown in Fig.~\ref{fig:cbelt}). Suppose that $a$ is the parameter of interest and $x$ is the measurement.

 \begin{figure}[h]
 \begin{centering}
 \includegraphics
[width=8 cm, angle=270, trim={50 0 0 0}, clip ] {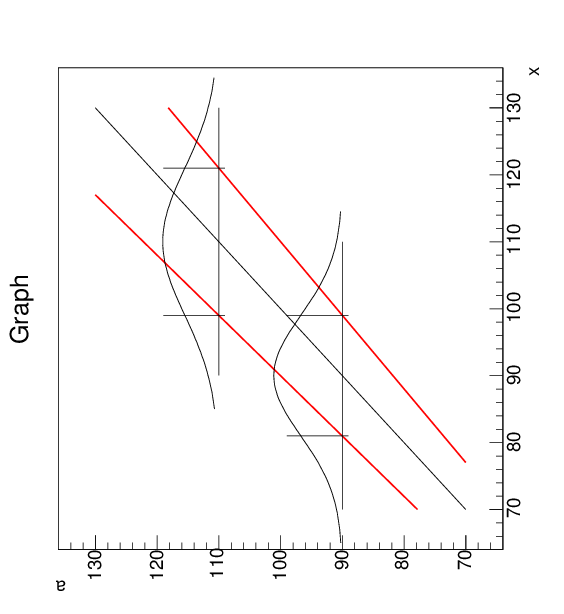}
\caption{\label{fig:cbelt} A confidence belt for a proportional Gaussian}
\end{centering}
\end{figure}

\begin{enumerate}
\item For each $a$, construct desired
 confidence interval 
 (here 68\% central).
\item The result $(x,a)$ lies inside the 
belt (the red lines), with 68\% confidence. 
\item Measure $x$.
 \item The result $(x,a)$ lies inside the 
 belt, with 68\% confidence.  And now we know $x$.
 \item Read off the belt limits  $a_+$ and $a_-$ at that $x$:  in this case they are 111.1, 90.9. 
 So we can report that  $a$ lies in [90.9,111.1] with 68\% confidence.
 \item Other choices for the confidence level value and for the strategy are available.
 \end{enumerate}
 
 This can be extended to the case of a Poisson distribution, Fig.~\ref{fig:confpois}. 
 
 \begin{figure}[h]
 \begin{centering}
 \includegraphics[width=8 cm, angle=270, trim={50 0 0 0}, clip ] {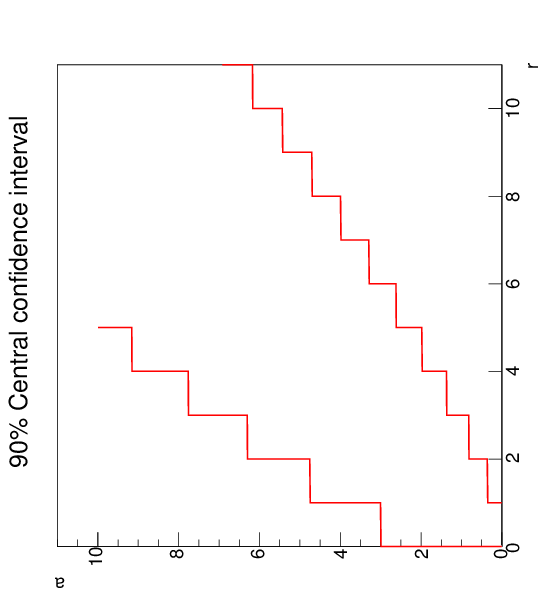}
\caption{\label{fig:confpois} A confidence belt for a Poisson}
\end{centering}
\end{figure}

The only difference is that the horizontal axis is discrete as the number observed, $x$, is integer.
In constructing the belt (horizontally) there will not in general be $x$ values available to give $\sum_{x_-}^{x_+}=CL$ and we call, again, on the `at least' in the definition and allow it to be $\sum_{x_-}^{x_+}\ge CL$.
 
Thus for a central 90\% confidence we require for each $a$ the largest integer  $x_{lo}$ and smallest $x_{hi}$
 for which 
$\sum_{x=0}^{x_{lo}-1} e^{-a}{a^x \over x!} \leq 0.05$
and 
$\sum_{x=x_{hi}+1}^{\infty} e^{-a}{a^x \over x!} \leq 0.05$.
For the second sum it is easier to calculate
$\sum_{x=0}^{x_{hi}} e^{-a}{a^x \over x!} \geq 0.95$\ .

Whatever  the value of $a$, the probability of the result falling in the belt is 90\%  {or more}.  We proceed as for the Gaussian.
 
\subsection{Coverage}

This is an appropriate point to introduce {\it coverage}:
the probability, given $a$, that the  statement `$a_{lo}\leq a \leq a_{hi}$' will be true.
Ideally this would be the same as the confidence level, however
it may (because of the `at least' clauses) 
 exceed it (`overcover'); this is allowed though in principle inefficient. It 
  should never be less (`undercover').

 For example: suppose we have a Poisson process with $a=3.5$ and we want a 90\% central limit.

 There is a probability $e^{-3.5}=3$\% of getting zero events, leading to $a_{+}=3.0$, which would be wrong as $3.0<3.5$\ . 
 
 Continuing in sequence, there is a probability $3.5 e^{-3.5}=$11\% of getting one event, leading to $a_{+}=4.7$, which would be  right.
 
 Right answers continue up to seven events (with probability $3.5^7 e^{-3.5}/7!=$4\% ): this gives a safely large value for $a_+$ and 
  $a_{-}=3.3$, which is right as $3.3<3.5$, though only just,  The next outcome, eight events (probability 2\%) gives $a_- = 4.0$ which is wrong, as are all subsequent results.

Adding up the probabilities for the outcomes 1 thru 7 that give a true answer totals 94\%,  so there is 4\% overcoverage.
 
 Note that coverage is a function of the true value of the parameter on which limits are being placed. 
 Values of $a$ other than 3.5 will give different coverage numbers---though all are over 90\%.
 
 \subsection{Upper limits}
 
 The one-sided upper limit---option 4 in the list above---gives us a way of quantifying the outcome of a null experiment.  `We saw nothing (or nothing that might not have been background), so we say $a\leq a_+$ at some confidence level'.

One simple and enlightening example occurs if you see no events, and there is no expected background. 
Now 
$P(0;2.996)=0.05$ and $2.996 \sim 3$. So if you see zero events, you can say with 95\% confidence that the 
true value is less than 3.0. You can then directly
use this to calculate a limit on the branching fraction, cross section, or whatever you're measuring.

 \subsection{Bayesian `credible intervals'}

A Bayesian has no problems saying `It will probably rain tomorrow'
or `The probability that $124.85 < M_H < 125.33$~GeV is 68\%'.  The 
downside, of course,  is that another Bayesian can say `It will probably not rain tomorrow' and 
`The probability that $124.85 < M_H < 125.33~GeV$ is 86\%' with equal validity and the two
cannot resolve their subjective difference in any objective way.

A Bayesian has a prior belief PDF $P(a)$ and defines a region $R$ such that $\int_R P(a)\, da = CL$.
There is the same ambiguity regarding choice of  content (68\%, 90\%, 95\%...) and strategy (central, symmetric, upper limit...). So Bayesian credible intervals look a lot like frequentist confidence intervals
even if their meaning is different.

 There are two happy coincidences.
 
The first is that 
Bayesian credible intervals on Gaussians, with a flat prior, are the same as Frequentist confidence intervals. If 
F quotes 68\% or 95\% or $\dots$ confidence intervals and B quotes  68\% or 95\% or $\dots$ credible interval,   
their results will agree.

The second is that  although the Frequentist Poisson upper limit is given by $\sum_{r=0}^{r=r_{data}} e^{-a_{hi}} a_{hi}^r / r!$ whereas the 
Bayesian Poisson flat prior upper limit is given by $\int_0^{a_{hi}}  e^{-a} a^{r_{data}} / r_{data}!\, da$,
 integration by parts of the Bayesian formula gives a series which is same as the Frequentist limit.
A Bayesian will also say : `I see zero events---the probability is 95\% that the true value is 3.0 or less.'
 This is (I think) a coincidence---it does not apply for lower limits. But it does avoid heated discussions as to which value to publish.
  
 \subsection{Limits in the presence of background}
 
 This is where  it gets tricky.
Typically an experiment may observe $N_D$ events, with an expected background $N_B$ and efficiency $\eta$, and wants to present results for  $N_S={N_D-N_B \over \eta}$.
Uncertainties in $\eta$ and $N_B$ are handled by profiling or marginalising.
The problem is that the 
{\it actual number} of background events is not $N_B$ but  Poisson in $N_B$.

So in a straightforward case, if you observe twelve events, with expected background 3.4 and $\eta=1$
it is obviously sensible to say $N_S=8.6$
(though the error is $\sqrt{12}$ not $\sqrt{8.6}$)
 
But suppose, with the same background, you see four events, three events or zero events.
Can you say $N_S=0.6$? or $-0.4$?  Or $-3.4$???
 
We will look at four methods of handling this, considering as an example the observation of three events with expected background 3.40 and wanting to present the 95\% CL upper limit on $N_S$.
 
\subsubsection{Method 1: Pure frequentist}

$N_D-N_B$ is an unbiased estimator of $N_S$ and its properties are known.
Quote the result.   Even if it is non-physical.

The argument for doing so
is that
this is needed for balance: if there is really no signal, approximately half of the experiments will give positive values and half negative.  
If the negative results are not published, but the positive ones are, the world average will be spuriously high.
For a 95\% confidence limit one accepts that 5\% of the results can be wrong.  This (unlikely) case is
clearly one of them.   So what?

 A counter-argument is that if
$N_D<N_B$, we {\it know} that the background has fluctuated downwards. But this cannot be incorporated 
into the formalism.

Anyway, the upper limit from 3 is 7.75, as $\sum_0^3 e^{-7.75}7.75^r/r! = 0.05$, and the 
95\% upper limit on $N_S=7.75-3.40=4.35$\ .

 \subsubsection{Method 2: Go Bayesian}

 Assign a uniform prior to $N_S$, for $N_S>0$, zero for $N_S<0$.
 The posterior is then just the likelihood, $P(N_S | N_D,N_B)=e^{-(N_S+N_B)}{(N_S+N_B)^{N_D} \over N_D!}$.
 The required limit is obtained from integrating $\int_0^{N_{hi}} P(N_S)\, dN_S = 0.95$
 where
 $P(N_S)\propto e^{-(N_s+3.40)}{(N_s+3.4)^3 \over 3!}$; this is illustrated in Fig.~\ref{fig:Bayeslimit}
 and the value of the limit is 
 5.21.
 
 \begin{figure}[h]
 \begin{centering}
 \includegraphics[ width=8 cm ] {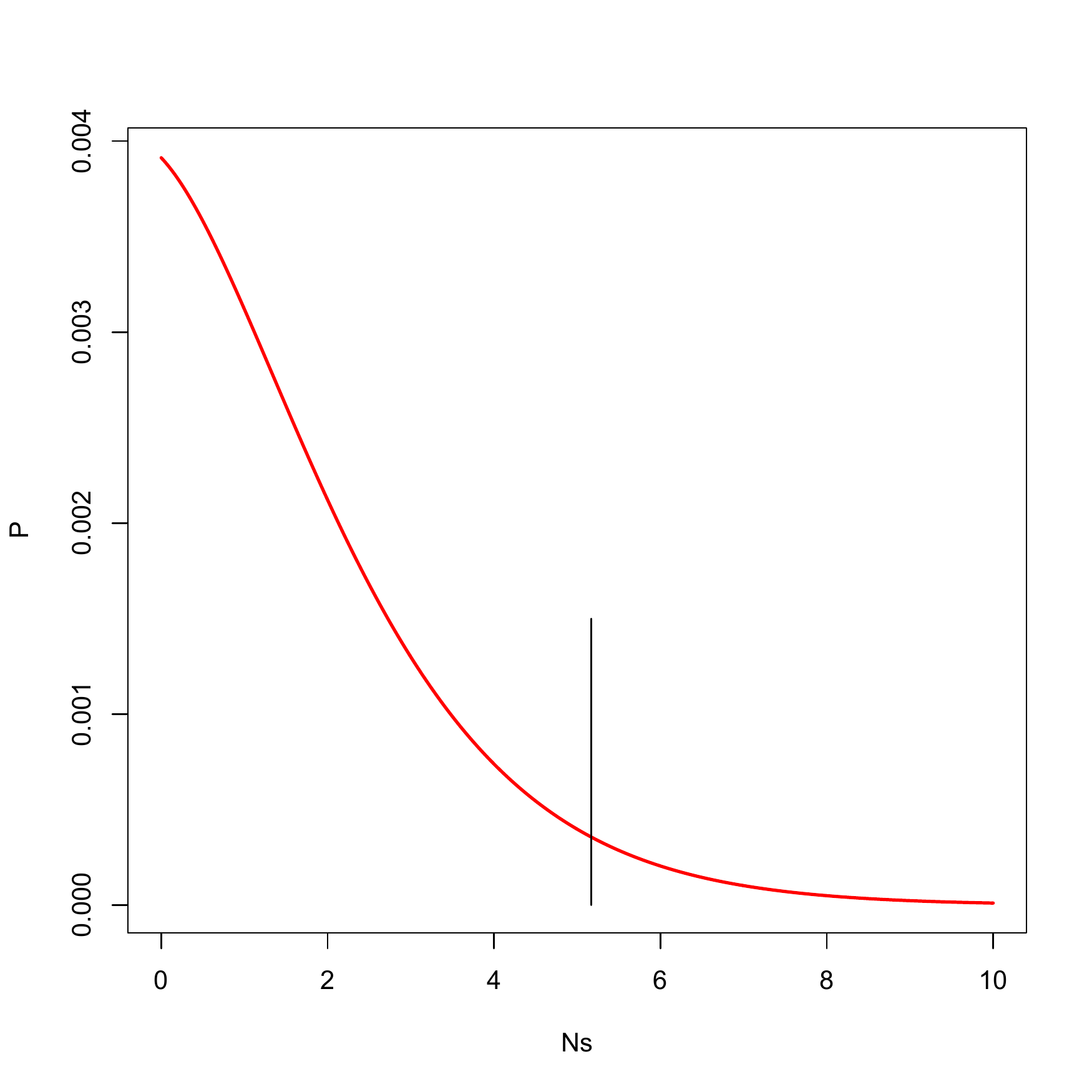}
 \caption{\label{fig:Bayeslimit} The Bayesian limit construction}
\end{centering}
\end{figure}

 \subsubsection{Method 3: Feldman-Cousins}

This---called `the unified approach' by Feldman and Cousins~\cite{FC}---takes a step backwards
and considers the ambiguity in the use of confidence belts.

In principle, if you decide to work at, say, 90\% confidence you may choose to use a 90\% central or a 90\% upper limit, and in either case  the probability of the result lying in the band is at least 90\%.
This is shown in Fig.~\ref{fig:FC1}.

In practice, if you happen to get a low result you would quote an upper limit,  but if you get a high result you would quote a central limit.
  This, which they call `flip-flopping',  is illustrated in the plot by a break shown here for $r=10$. 
  \begin{figure}[h]
 \begin{centering}
 \includegraphics[ width=8 cm, angle=270 ] {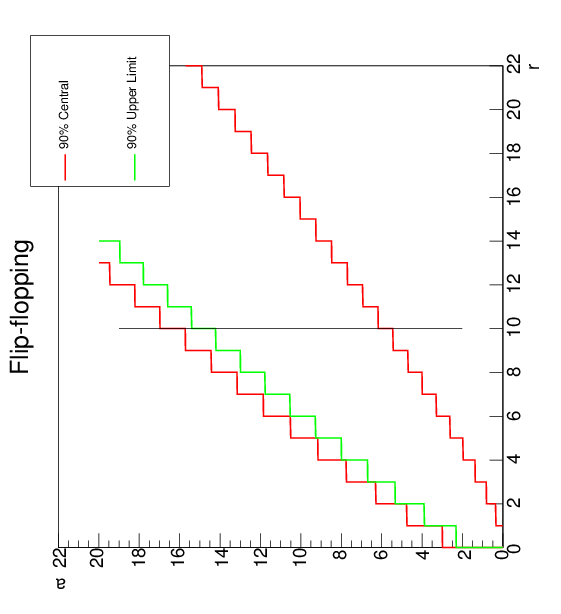}
\caption{\label{fig:FC1} The flip-flopping problem}
\end{centering}
\end{figure}

Now the confidence belt is the green one for $r< 10$ and the red one for $r\geq 10$. The
probability of lying in the band is no longer  90\%! 
Flip-flopping invalidates the Frequentist construction, leading to undercoverage.  

 They show how to avoid this. You draw the plot slightly differently:
 $r \equiv N_D$ is still the horizontal variable,  but as the vertical variable you use  $N_S$. 
  (This means a different plot for any different $N_B$, whereas the previous Poisson plot is universal, but this is not a problem.)
 This is to be filled using  $P(r;N_s)=e^{-(N_s+N_B)}{(N_S+N_B)^r \over r!}$\ .

For each $N_S$ you  define a region $R$ such that $\sum_{r\epsilon R}P(r;N_s) \geq 90\%$.  
You have a choice of strategy that goes beyond `central' or `upper limit': one  
plausible suggestion would be to
 rank $r$ by probability and take them in order until the desired total probability content is achieved  (which would, incidentally, give the shortest interval).
However this has the drawback that outcomes with $r < N_B$  will have small probabilities and be excluded for all $N_S$, so that, if such a result does occur, one cannot say anything constructive, just `This was unlikely'. 

An improved form of this  suggestion is that for each $N_S$, considering  each $r$ you compare $P(r;N_S)$ with the largest possible value obtained by varying $N_S$. This is easier than it sounds because this highest value is  either at $N_S=r-N_B$ (if $r\geq N_B$) or $N_S=0$ (if $r\leq N_B$ ).
Rank on the ratio $P(r;N_S)/P(r;N^{best}_S)$ and again take them in order till their sum gives the  desired probability.

This gives a band as shown in Fig.~\ref{fig:FC2}, which has $N_B=3.4$. You can see that 
`flip-flopping' occurs naturally: for small values of $r$ one just has an upper limit, whereas for larger values, above $r=7$, one obtains a lower limit as well. Yet there is a single band, and the coverage is
correct (i.e. it does not undercover).
In the case we are considering, $r=3$, just an upper limit is given, at $4.86$. 
  \begin{figure}[h]
 \begin{centering}
 \includegraphics[ width=8 cm, angle=270 ] {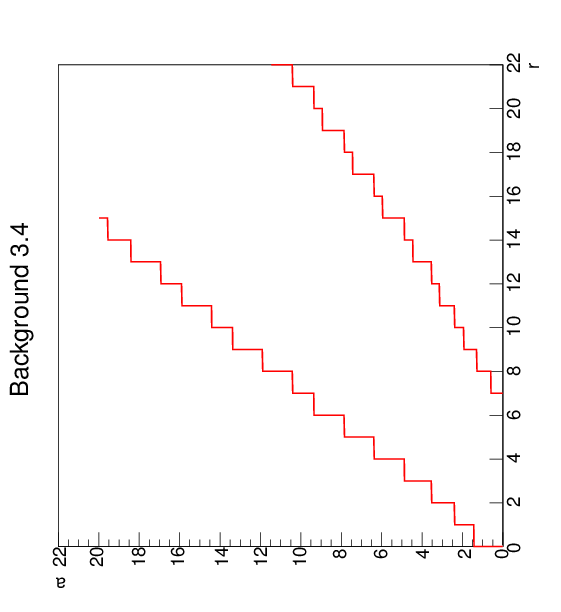}
\caption{\label{fig:FC2} A Feldman-Cousins confidence band}
\end{centering}
\end{figure}
 
Like other good ideas, this has not found universal favour. Two arguments are raised against the method.

First, that it  deprives the physicist of the choice of whether to publish  an upper limit or a range. 
It could be embarrassing if you
look for something weird and are `forced' to publish a non-zero result. 
But this is actually the point, and in such cases one can always explain
that the limits should not be taken as implying that the quantity actually is nonzero.

Secondly, if two experiments with different $N_B$ get the same small $N_D$, the one with the higher $N_B$ will quote a smaller limit on $N_S$.   The worse experiment gets the better result, which is clearly unfair!
But this is not comparing like with like: for a `bad' experiment with large background to get a small number of events is much less likely than it is for a `good' low background experiment.
 
 \subsubsection {Method 4: $CL_s$}

This is a modification of the standard frequentist approach to include the 
fact, as mentioned above, that a small observed signal implies a downward 
fluctuation in background~\cite{Read}. Although presented here using just numbers of events, the method is usually extended to use the full likelihood of the result, as will be discussed in Section~\ref{subsection:Extension}.

 \begin{figure}[h]
 \begin{centering}
 \includegraphics[ width=8 cm, angle=270 ] {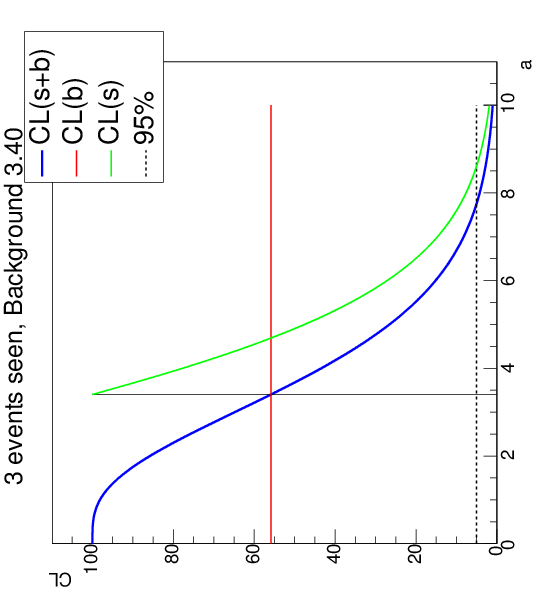}
\caption{\label{fig:CLS} The $CL_s$ construction}
\end{centering}
\end{figure}

Denote the (strict frequentist) 
probability of getting a result this small (or less) from $s+b$ events as 
  $CL_{s+b}$, and the equivalent probability from pure background as 
  $CL_b$ (so 
 $CL_b=CL_{s+b}$ for $s=0$).
 Then introduce
 
\begin{equation}
 CL_s={CL_{s+b} \over CL_b}
 \label{eq:CLs}
 \quad.
 \end{equation}

Looking at  Fig.~\ref{fig:CLS}, the  $CL_{s+b}$ curve shows that if $s+b$ is small then the probability of getting three events or less is high, near 100\%. As $s+b$ increases this probability falls, and at $s+b=7.75$ 
the probability of only getting three events or less is only 5\%. This, after subtraction of $b=3.4$, gives the strict frequentist value.

The point $s+b=3.4$ corresponds to $s=0$,  at which the probability $CL_b$ is  56\%  As $s$ must be non-negative, one can argue that everything to the left of that is unmeaningful.  So one attempts to 
incorporate this by renormalizing the (blue) $CL_{s+b}$ curve to have a maximum of 100\% in the 
physically sensible region, dividing it by 0.56 to get the (green) $CL_s$ curve.
This is treated in the same way as the $CL_{s+b}$ curve, reading off the point at $s+b=8.61$ where it falls to 5\%.   This is a limit on $s+b$ so we subtract 3.4 to get the limit on $s$ as 5.21.
This is  larger than the strict frequentist limit: the method over-covers (which, as we have seen, is allowed if not encouraged)
  and is, in this respect `conservative'\footnote{`Conservative' is a misleading word. It is used by people 
  describing their analyses to 
  imply safety and caution, whereas it usually entails cowardice and  sloppy thinking.}. This is the same value as the Bayesian Method 2, as it makes the same assumptions. 
  
  $CL_s$ is not frequentist, just `frequentist inspired'. In terms of statistics there is perhaps little  in its favour. But it has an intuitive appeal, and is widely used.
  
  \subsubsection{Summary so far}

Given three observed events, and an expected background of 3.4 events, what is
the 95\% upper limit on the `true' number of events?
Possible answers are shown in table~\ref{tab:summary}.

 \begin{table}[h]
\begin{center}
\begin{tabular}{|c|c|}
\hline
Strict  Frequentist & 4.35 \\
Bayesian (uniform prior) & 5.21 \\
Feldman-Cousins &  4.86   \\
$CL_s$ & 5.21 \\
\hline
\end{tabular}
\end{center}
\caption{\label{tab:summary} Upper limits from different methods}
\end{table}

Which is `right'? Take your pick!
All are correct.  (Well, not wrong.). The golden rule is  to say what you are doing, and if possible give the raw numbers. 
 
  \subsubsection{Extension: not just  counting numbers}
  \label{subsection:Extension}
 \begin{figure}[h]
 \begin{centering}
\centerline{\includegraphics[width=8 cm, angle=270, trim={0 0 0 0},clip]{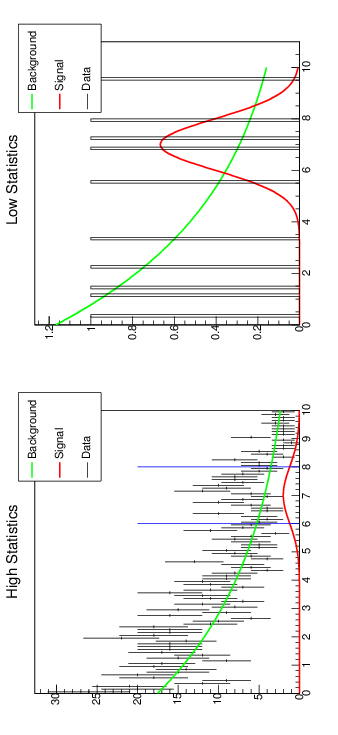}}
\caption{\label{fig:beyondsimple} Just counting numbers may not give the full information}
\end{centering}
\end{figure}

These examples have used 
simple counting experiments. But a simple number does not (usually) exploit the full information.

Consider the illustration in Fig.~\ref{fig:beyondsimple}.  One is searching for (or putting an upper limit on) some broad resonance around 7~GeV. One could count the number of events inside some window
(perhaps 6 to 8~GeV?) and subtract the estimated background.  This might work with high statistics, as in the left, but would be pretty useless with small numbers, as in the right. It is clearly not optimal 
just to count an event as `in', whether it is at  7.0 or 7.9, and to treat an event as `out', if it is at 8.1 or 
10.1.

It is better to calculate the 
Likelihood $\ln L_{s+b}=\sum_i \ln{N_s S(x_i)+N_b B(x_i)} \quad;\quad \ln{ L_b}=\sum_i \ln{N_b B(x_i)}$.
Then, for example using $CL_s$, you can work with  $L_{s+b}/L_b$, or  $-2 \ln{(L_{s+b}/L_b)}$.
The  confidence/probability  quantities can be found from simulations, or sometimes from data.

 \subsubsection{Extension: From numbers to masses}

Limits on numbers of events can readily be  translated into limits on branching ratios,
$BR={N_s \over N_{total}}$,
or limits on cross sections,
$\sigma={N_s \over \int {\cal L} dt}$\ .

These may translate to limits on other, theory, parameters.

In the Higgs search (to take an example) the cross section depends on the mass,  $M_H$---and so does the detection efficiency---which may require changing strategy (hence different backgrounds). This leads to the need
 to basically repeat the analysis for all (of many) $M_H$ values. This can be presented in two ways.

  \begin{figure}
 \begin{centering}
 \includegraphics[width=8 cm]{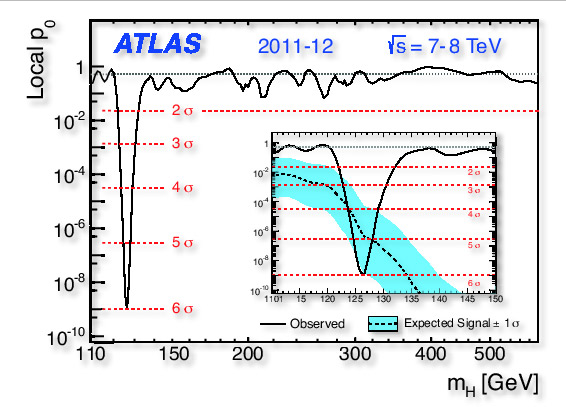}
\caption{\label{fig:significanceplot} Significance plot for the Higgs search}
\end{centering}
\end{figure}

  The first is shown in Fig.~\ref{fig:significanceplot}, taken from Ref.~\cite{ATLAS1}. For each $M_H$ (or whatever is being studied) you search for a signal and plot the $CL_s$ (or whatever limit method you prefer) significance  
   in a {\it Significance Plot}. 
  Small values indicate that it is  unlikely to get a signal this large just from background.
  
  One often also plots the  expected (from MC) significance, assuming the signal hypothesis is true. This is a measure of a  `good experiment'. In this case there is a discovery level 
  drop at $M_H \approx 125$~GeV, which exceeds the expected significance, though not by much: ATLAS were lucky but not incredibly lucky.
  
  The second method is---for some reason---known as the green-and-yellow plot.
  This is basically the same data, but fixing $CL$ at a chosen value: in Fig.~\ref{fig:greenandyellow} it is 95\%.
 You find the  limit on signal strength, at this confidence level,  and interpret it as 
 a limit on the cross section $\sigma / \sigma_{SM}$.
  Again, as well as plotting the  actual data  one also plots the expected (from MC) limit, with variations.
 If there is no signal, 68\% of experiments should give results in the green band, 95\% in the yellow band. 
   
  \begin{figure}[h]
 \begin{centering}
 \includegraphics[width=8 cm]{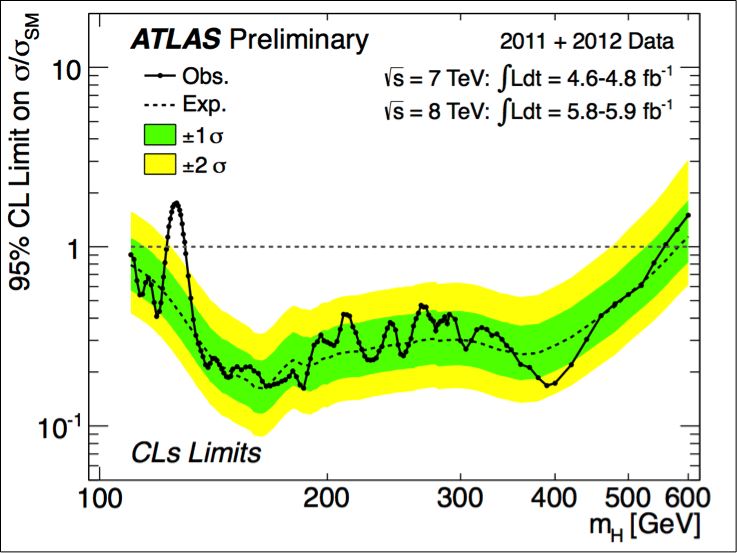}
 \caption{\label{fig:greenandyellow} Green and yellow plot showing the Higgs discovery}
\end{centering}
\end{figure}

So this figure shows the experimental result as a black line.  Around 125~GeV the 95\% upper limit 
is more than the Standard Model prediction indicating a discovery. There are peaks between 200 and 300~GeV, but they do not approach the SM value, indicating that they are just fluctuations. The value rises at 600~GeV, but the green (and yellow) bands rise also, showing that the experiment is not
sensitive for such high masses: basically it sees nothing but would expect to see nothing.

  \section{Making a discovery}
  
  We now turn from setting limits, to say what you did not see,
   to the more exciting prospect of making a discovery.

\label{sec:discovery}

Remembering hypothesis testing, in claiming a discovery  you have to  show that your data can't be explained without it.
This is 
quantified by the $p-$value: the probability of getting a result this extreme (or worse) under the null hypothesis/Standard Model. 
(This is {\it not} `The probability that the Standard Model is correct', but it seems impossible for journalists
to understand the difference.)
 
 Some journals (particularly in psychology) refuse to publish papers giving $p-$values.
If you do lots of studies, some will have low $p-$values  (5\% below 0.05 etc.). 
The danger is that these get published, but the unsuccessful ones are binned.

Is $p$ like the significance $\alpha$? Yes and no.  The formula is the same, but $\alpha$ is a property of the test, computed before you see the data.
$p$ is a property of the data. 
 
 \subsection{Sigma language}

 The probability ($p-$value) is often
  translated into Gaussian-like language: the probability of a result more than 3$\sigma$ from the mean is 0.27\% so a $p-$value of 0.0027 is a `3 $\sigma$ effect' (or 0.0013 depending on
  whether one takes the 1-tailed or 2-tailed option. Both are used.)
In reporting a result with a significance of  `so many $\sigma$' there is no actual 
$\sigma$ involved: it is just a translation to give a better feel for the size of the probability.

By convention, 3 sigma, $p= 0.0013$  is reported as `Evidence for' whereas a full  
5 sigma\\ $p=0.0000003$ is required for  `discovery of'.

\subsection{The look-elsewhere effect}
  
 You may think that the requirement for 5 $\sigma$ is excessively cautious.
 Its justification comes from history---too many 3- and 4- sigma `signals' have gone away when more data was taken.
 
 This is partly explained by the `look-elsewhere effect'.  How many peaks can you see in the
 data in Fig.~\ref{fig:LEE}?

  \begin{figure}[h]
 \begin{centering}
 \centerline{ \includegraphics[width=6 cm,angle=270, trim={260 0 0 0},clip]{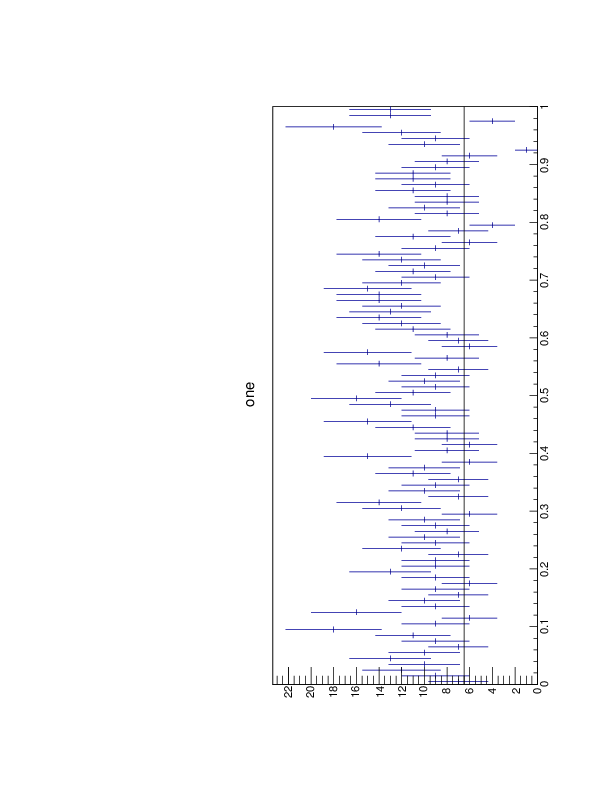}}
 \caption{\label{fig:LEE} How many peaks are in this data?}
\end{centering}
\end{figure}

  The answer is that there are none. The data is in fact purely random and flat. But the human eye is very good at seeing features.
 
 With 100 bins, a $p-$value below 1\% is pretty likely.
 This can be factored in, to some extent, using pseudo-experiments, but this  does
 not allow for the sheer number of plots being produced by 
 hard-working physicists looking for something.  Hence the need for caution.
  
 This is not just ancient history. ATLAS and CMS recently observed a signal in the $\gamma \gamma$ mass around 750~GeV, with a significance of
  $3.9 \sigma$ (ATLAS) and $3.4 \sigma$  (CMS), which went away when more data was taken.

 \subsection{Blind analysis}
  
  It is said\footnote{This story is certainly not historically accurate, but it's still a good story (\textit{quoteinvestigator.com}: \url{https://quoteinvestigator.com/2014/06/22/chip-away/}).}  that when Michaelangelo was asked how he created his masterpiece sculpture `David' 
  he replied
 `It was easy---all I did was get a block of marble and chip away everything that didn't look like David'.
 Such creativity may be good for sculpture, but it's bad for physics. 
 If you take  your data and devise cuts to remove all the events that don't look like the signal you want to see, then whatever is left 
 at the end will look like that signal. 
  
  Many/most analyses are now done `blind'. 
  Cuts are devised using Monte Carlo and/or non-signal data.
 You only `open the box' once the cuts are fixed.  Most collaborations have a formal procedure for doing this.
 
 This may seem a tedious imposition, but we have learnt the hard way that it avoids embarrassing  mistakes.
   
  \section{Conclusions}
  
Statistics is a tool for doing physics.
Good physicists understand their tools.  Don't just follow without understanding, but 
read books and conference proceedings, go to seminars, talk to people, experiment with the data, and understand
what you are doing.
Then you will succeed.
And you will have a great time!

\begin{flushleft}

\end{flushleft}
\end{document}